\documentclass[aps,prb,twocolumn,superscriptaddress,10pt]{revtex4-1}

\usepackage{graphicx}
\usepackage{xcolor}
\usepackage{amsmath}

\begin{document}


\title{Full electrostatic control of quantum interference \\ in an extended trenched Josephson junction}



\author{Stefano Guiducci}
\thanks{deceased 19 October 2018}
\affiliation{NEST, Istituto Nanoscienze-CNR and Scuola Normale Superiore, Piazza San Silvestro 12, 56127 Pisa, Italy}

\author{Matteo Carrega}
\affiliation{NEST, Istituto Nanoscienze-CNR and Scuola Normale Superiore, Piazza San Silvestro 12, 56127 Pisa, Italy}

\author{Fabio Taddei}
\affiliation{NEST, Istituto Nanoscienze-CNR and Scuola Normale Superiore, Piazza San Silvestro 12, 56127 Pisa, Italy}

\author{Giorgio Biasiol}
\affiliation{IOM CNR, Laboratorio TASC, Area Science Park, 34149 Trieste, Italy}

\author{Herv\'e Courtois}
\affiliation{Univ.~Grenoble Alpes, CNRS, Institut N\'eel, 25 avenue des martyrs, Grenoble, France}

\author{Fabio Beltram}
\affiliation{NEST, Istituto Nanoscienze-CNR and Scuola Normale Superiore, Piazza San Silvestro 12, 56127 Pisa, Italy}

\author{Stefan Heun}
\email{stefan.heun@nano.cnr.it}
\affiliation{NEST, Istituto Nanoscienze-CNR and Scuola Normale Superiore, Piazza San Silvestro 12, 56127 Pisa, Italy}


\date{\today}

\begin{abstract}
Hybrid semiconductor/superconductor devices constitute an important platform for a wide range of applications, from quantum computing to topological--state--based architectures. Here, we demonstrate full modulation of the interference pattern in a superconducting interference device with two parallel islands of ballistic InAs quantum wells separated by a trench, by acting independently on two side--gates. This so far unexplored geometry enables us to tune the device with high precision from a SQUID-like to a Fraunhofer-like behavior simply by electrostatic gating, without the need for an additional in-plane magnetic field. These measurements are successfully analyzed within a theoretical model of an extended tunnel Josephson junction, taking into account the focusing factor of the setup. The impact of these results on the design of novel devices is discussed.
\end{abstract}


\maketitle

\section{\label{sec:Introduction}Introduction}

In the last decades, much attention has been devoted to the development of quantum technologies based on hybrid systems with superconducting elements. Different applications relying on the celebrated Josephson effect \cite{Barone1982,Tinkham1996}, are already available, such as magnetometers \cite{Ronzani2014,Jabdaraghi2018}, single-photon detectors \cite{Govenius2016}, and quantum computing architectures \cite{Wendin2017}. Recently, the interest in hybrid Josephson junctions (JJs) was also fueled by the intriguing possibility of hosting new topological states of matter \cite{Mourik2012,Beenakker2013,Mong2014,Albrecht2016}. Among all proposals, a promising route toward the realization of new topological states relies on the coupling between semiconductors with strong spin-orbit interactions and superconductors \cite{Mourik2012,Beenakker2013}.

JJs are also widely used to study the interplay between superconductivity, spin-orbit interactions, and external fields in hybrid devices. Important information is encoded in the behavior of the Josephson supercurrent $I_c$, which shows characteristic interference patterns, i.e.~modulations as a function of an external out-of-plane magnetic field. By measuring these, one can thus study quantities such as local magnetic profiles or supercurrent densities within the hybrid junction \cite{Ishikawa1999, Silaev2017}.

Josephson-based interferometers \cite{Nakamura1999,Friedman2000,Yu2002} require very good --- low--resistance --- normal/su\-per\-con\-duc\-tor (N/S) contacts, yielding robust proximity effect \cite{Belzig1999,Courtois1999a} and a sufficiently large electron elastic mean free path in the N region \cite{Amado2014}. Semiconducting quantum wells of In$_{x}$Ga$_{1-x}$As (with molar fraction $x\geq 0.75$) with superconducting Niobium contacts were used in the past to meet these constraints \cite{Fornieri2013,Drachmann2017}. However, pure InAs quantum wells represent the ideal choice for building hybrid devices, owing to the lack of a Schottky barrier at the interface with the metal, combined with its small effective mass and large spin-orbit coupling \cite{Takayanagi1995,Shabani2016,Kjaergaard2016,Kjaergaard2017,Goffman2017,Sestoft2018}. 

Recently, epitaxial Al/InAs heterostructures were realized that showed an exceptionally transparent superconductor-semiconductor interface, resulting in almost ideal Andreev reflections \cite{Goffman2017,Casparis2017}. The magnetic-field dependence of the interference pattern of the critical current in epitaxial Al/InAs/Al junctions was reported, both with a perpendicular field and a separately-controlled in-plane field \cite{Suominen2017}. By tuning the latter, it was shown that a crossover in the perpendicular--field interference pattern appears with increasing in--plane field, from a Fraunhofer--like shape (characteristic of an extended junction) towards one resembling that of a superconducting quantum interference device (SQUID). This approach, however, requires the tuning of an in-plane magnetic field and is therefore not scalable.

A major advantage offered by 2D semiconductor hybrids is the possibility to electrostatically tailor and manipulate superconducting properties by means of additional gates \cite{Amado2013,Paajaste2015,Kjaergaard2016,Guiducci2018,Ke2019}. Indeed, magnetic interference patterns of supercurrent can be tailored by shrinking electrostatically the width of the normal region via lateral gates, as shown for single JJs \cite{Amado2013,Paajaste2015,Kjaergaard2016,Guiducci2018,Seredinski2019} and, more recently, in the SQUID geometry \cite{Monteiro2017,Thompson2017}.

In this work, we present a new JJ device in which a high-quality InAs quantum well is placed between two Nb superconducting contacts. In addition, a hole in the central region of the 2DEG has been introduced, forming an extended trenched Josephson junction with two closely placed parallel device arms. By independently acting on two side--gates, we achieve full modulation and control of the interference patterns of the device as a function of an out-of-plane magnetic field, continuously modifying the interference period of the critical current and moving from a Fraunhofer-like pattern to a monotonically decaying one, thus exploring both the wide and narrow-junction limits. Our results show, for the first time, that within a single device it is possible to tune supercurrents at will by electrostatic gating from a SQUID--like to a Fraunhofer--like behavior. Experimental data are explained using a theoretical model, corroborating a clear and consistent physical picture. 

\section{\label{sec:ExperimentalResults}Experimental Results}

\begin{figure}[t]
   \includegraphics[width=\columnwidth]{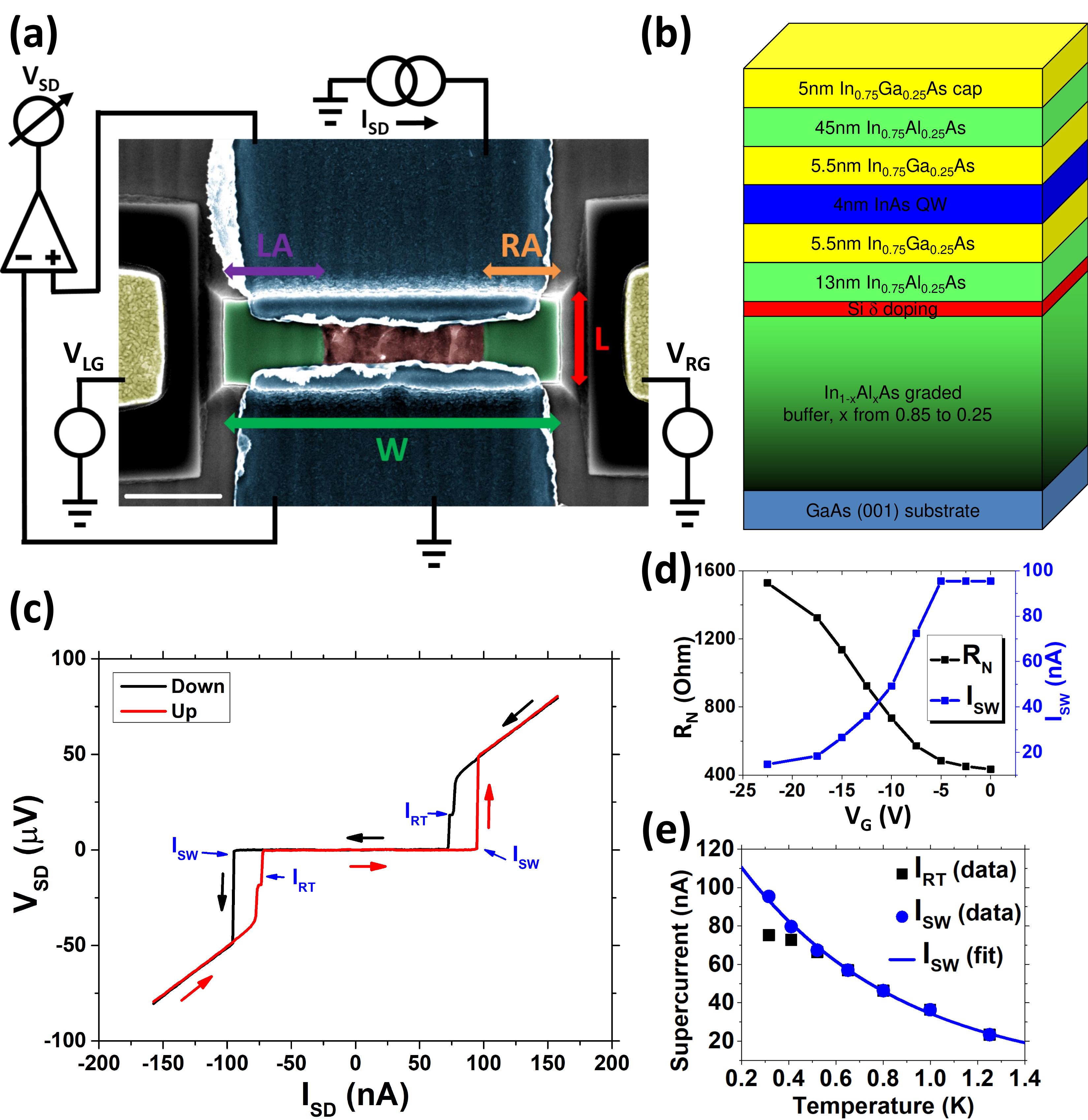}
   \caption{\label{fig:1}(a) Pseudo-color scanning electron micrograph of the measured device (top view): the superconductor is shown in blue, the N region in green, the hole in the 2DEG in red, and the side--gates in yellow. Standard current-bias four-wire setup is sketched. The resistance of each line is omitted (2.1~k$\Omega$, mostly originating from the RC filters). (b) Schematic drawing of the heterostructure: the 2DEG is confined in the InAs region (blue); the Si-doped region is marked red; below there is the buffer region in order to accommodate the lattice mismatch between InAs and GaAs. (c) $I_{SD}-V_{SD}$ curves of the measured device. The red (black) arrows show the direction of the up (down) sweep. From this measurement, a switching current $I_{SW} = (95 \pm 1)$~nA and a retrapping current $I_{RT} = (77 \pm 1)$~nA are deduced. $T = 315$~mK, $B = -0.21$~mT, $V_{LG} = V_{RG} = 0$~V. (d) Maximum switching current $I_{SW}$ (blue) and normal resistance $R_N$ (black) as a function of gate voltage $V_{G}$ ($V_{G} = V_{LG} = V_{RG}$) at $T = 315$~mK. The side--gates offer a strong modulation of the supercurrent by squeezing simultaneously the two arms of the hybrid device. (e) Maximum switching current $I_{SW}$ (blue) and maximum retrapping current $I_{RT}$ (black) as a function of temperature, for $V_{G}=0$~V.}
\end{figure}

Figure~\ref{fig:1}(a) shows the device investigated in this work. The superconducting Nb leads are colored in blue, the normal weak-link in green, the removed central region of the mesa in red, and the side--gates in yellow. Figure~\ref{fig:1}(b) shows a schematic drawing of the heterostructure used for the N region. It consists of an InAs quantum well, which guarantees exceptionally transparent superconductor/semiconductor interfaces. The inter-electrode spacing between the two superconductors, i.e.~the length of the N region, is $L = 0.9$~$\mu$m.  The width of the {weak-link} is $W = 3.6$~$\mu$m, while the width of the left arm (LA) and right arm (RA) are $1.0$~$\mu$m and $0.75$~$\mu$m, respectively. Standard transport characterization (see Methods) yields a mean free path of $\ell = 2$~$\mu$m for the N region, greater than the junction length $L$.

Additional voltage probes (not shown in Fig.~\ref{fig:1}(a)) are present on the sample in order to measure the voltage drop across the Nb superconductive leads. We have measured a Nb critical field $H_{c2}$ of 2.7~T (at $T = 315$~mK) and a critical temperature $T_c \simeq 8.1$~K, resulting in an estimated BCS Nb gap $\Delta_{Nb} = 1.76 k_B T_c \simeq 1.2$~meV \cite{Fornieri2013a,Guiducci2014}. With $v_{F,Nb} = 1.37 \times 10^6$~m/s the Fermi velocity in Nb \cite{Ashcroft1976}, from the gap we extract a BCS coherence length $\xi = \frac{1}{\pi} \frac{\hbar v_{F, Nb}}{\Delta_{Nb}} = (239 \pm 4)$~nm~$\ll L$, which indicates that the device operates in the long-junction regime \cite{Mur1996,Grosso2000,Fornieri2013a}.

The SNS device was characterized in a filtered He-3 cryostat down to 315~mK. The structure was biased by a current $I_{SD}$, while the voltage drop $V_{SD}$ across the mesa was registered via a room-temperature differential preamplifier [see Fig.~\ref{fig:1}(a)]. Figure~\ref{fig:1}(c) shows a typical $V_{SD}$ vs.~$I_{SD}$ curve obtained in a four-wire setup. Current sweeps in opposite directions (black and red) show a hysteretic behavior, commonly understood as Joule heating of the 2DEG in the dissipative regime \cite{Schapers2001,Courtois2008,Fornieri2013a}. The $I-V$ curve was measured in the condition of maximum supercurrent, at $T = 315$~mK, zero gate voltage ($V_{LG}=V_{RG}=0$~V), and $B = -0.21$~mT (a small magnetic field is used to compensate for the residual magnetization in the cryostat).

\begin{figure*}[t]
   \includegraphics[width=\textwidth]{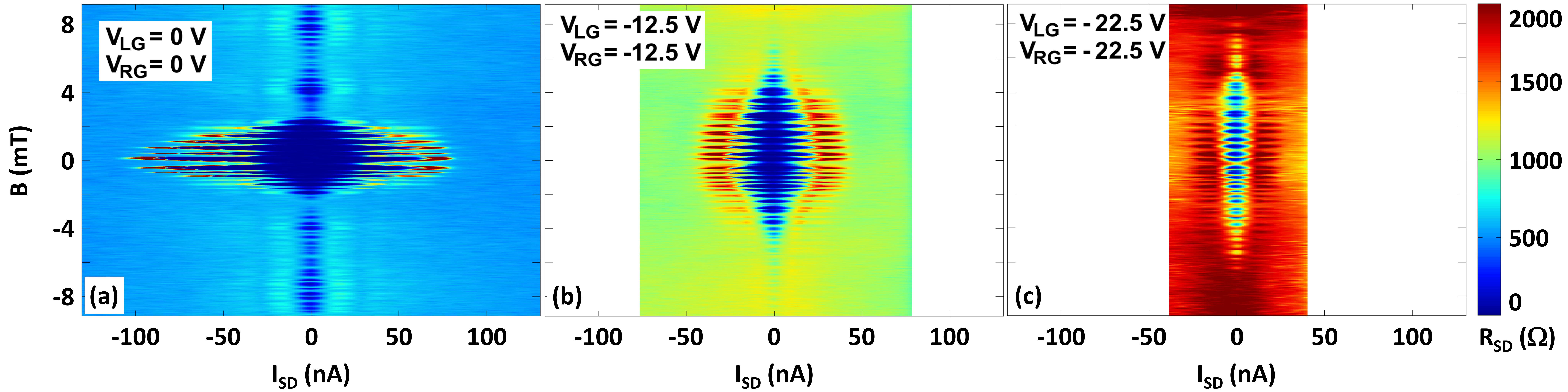}
   \caption{\label{fig:2} Differential resistance of the device as a function of the injected source-drain current and magnetic field in three different configurations of symmetric gate voltage: (a) $V_{LG} = V_{RG} = 0$~V; (b) $V_{LG} = V_{RG} = -12.5$~V; (c) $V_{LG} = V_{RG} = -22.5$~V. The dark blue region is the Josephson regime. As the gate voltage becomes more negative, the magnetic interference pattern changes towards a SQUID-like. $T = 315$~mK.}
\end{figure*}

Figure~\ref{fig:1}(d) shows switching current ($I_{SW}$) and normal resistance ($R_N$) of the device versus gate voltage $V_{G}$ ($V_{G} = V_{LG} = V_{RG}$). $R_N$ was measured at high bias ($|V_{SD}| > 2\Delta_{Nb}/e$) where Andreev reflections are completely suppressed. The product $I_{SW} R_N$, known as the critical voltage $V_c$ \cite{Marsh1994}, reaches a maximum value of 46~$\mu$V at $V_{G} = -5$~V and then rapidly drops for more negative $V_{G}$ due to the strong decrease of $I_{SW}$. From $I-V$ curves at high bias and $V_{G}=0$~V, we extract the excess current $I_{\text{exc}} =330 \pm 10 ~\text{nA}$. Assuming that the resistance of the junction is dominated by the interfaces, this value is used to calculate the dimensionless barrier strength $Z$ from the OBTK model \cite{Blonder1982,Octavio1983,Flensberg1988}. This gives $Z = 1 \pm 0.04$, or equivalently, an interface transparency $\tau = 0.50 \pm 0.02$, with $\tau = 1/(1 + Z^2)$.

The temperature evolution of switching current $I_{SW}$ (blue) and retrapping current $I_{RT}$ (black) is reported in Fig.~\ref{fig:1}(e). Both values decrease with temperature, and above 500~mK the hysteresis vanishes. In the following, we shall use switching current and critical current as synonyms, \mbox{i.e.} $I_{SW}=I_c$. The figure also shows the fitted exponential decay function $I_{SW} (T) = I_0 \exp{(- T/T_0)}$ (blue). The parameters that best fit the data are: $I_0 = (148 \pm 4)$~nA and $T_0 = (0.69 \pm 0.02)$~K. For long ballistic junctions, the critical current is expected to follow $I_c \propto \exp(-L/\xi_n)$ \cite{Kulik1970,Ishii1970,Bardeen1972,Likharev1979,Chrestin1994,Mur1996,Schapers2001,Dubos2001,Giazotto2004}, with the thermal coherence length in the clean limit $\xi_n(T) = \hbar v_{F, 2DEG} / (2 \pi k_B T)$ \cite{Chrestin1994,Marsh1994,Giazotto2004}, where $v_{F, 2DEG} = (\hbar / m) \sqrt{2 \pi n}$ is the Fermi velocity in the 2DEG \cite{Chrestin1994}. With $m = 0.03 m_{e}$ and $n = 3.74 \times 10^{11}$~cm$^{-2}$ (see Methods) we obtain here $v_{F, 2DEG} = 5.92 \times 10^5$~m/s. The fitted temperature scale $T_0 = (0.69 \pm 0.02)$~K agrees very well with the calculated value $T_0 = \hbar v_{F, 2DEG} / (2 \pi k_B L) = 0.72$~K. The good agreement with an exponential decay law confirms that the transport in the N region is ballistic \cite{Mur1996,Schapers2001,Dubos2001}, consistently with the fact that the measured mean free path is larger than the device length, $\ell > L$ \cite{Fornieri2013a,Fornieri2013,Amado2014}.

Figure~\ref{fig:2}(a) reports the differential resistance $R_{SD}=dV_{SD}/dI_{SD}$ of the JJ as a function of applied magnetic field and source-drain current. The dark blue region in the plot represents the Josephson regime, i.e. $R_{SD} \approx 0$~$\Omega$. A strong modulation of the differential resistance is observed as a function of the applied magnetic field. Two contributions are clearly visible: a slowly varying ``envelope" component and a fast-varying one that modulates the former.

\begin{figure*}[tbh]
   \includegraphics[width=\textwidth]{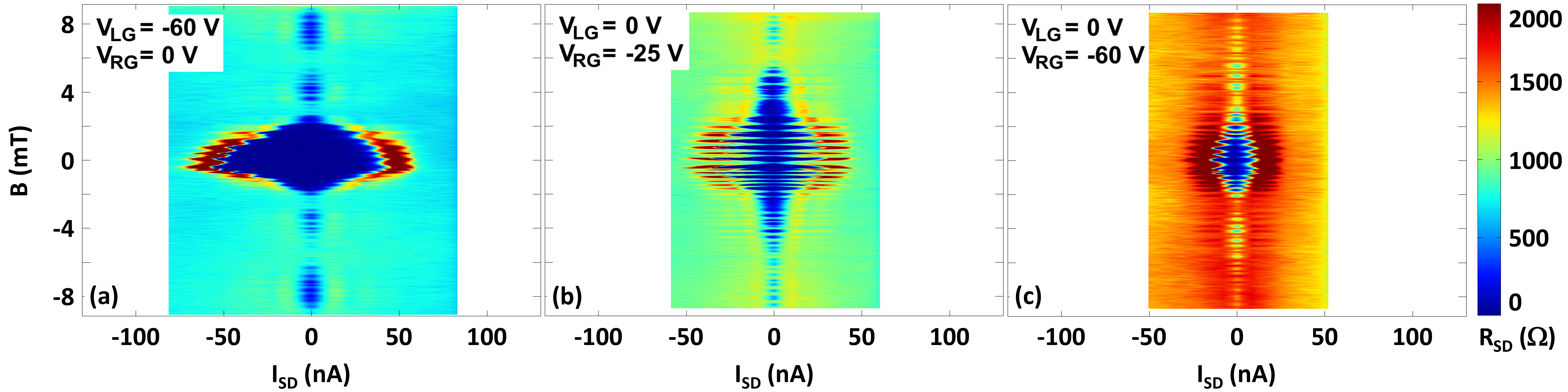}
   \caption{\label{fig:3} Differential resistance of the device as a function of the injected source-drain current and magnetic field in three different configurations of asymmetric gate voltage: (a) $V_{LG} = -60$~V and $V_{RG} = 0$~V; (b) $V_{LG} = 0$~V and $V_{RG} = -25$~V; (c) $V_{LG} = 0$~V and $V_{RG} = -60$~V. The center of the pattern in (b) presents a slight shift of $-0.2$~mT with respect to $B=0$, consistent with the residual magnetization of the cryostat. $T = 315$~mK.}	
\end{figure*}

An analysis of the spacing between consecutive minima of the ``envelope" component provides the periodicity of the slow contribution: $\Delta B_{slow} = (2.89 \pm 0.25)$~mT. With $\Phi_0=h/(2 e)$ the flux quantum, we calculate the corresponding characteristic area as \cite{Tinkham1996,Suominen2017} $A_{slow} = \Phi_0 / \Delta B_{slow} = (0.72 \pm 0.06)$~$\mu$m$^2$, which is in between the areas of the two arms of the device: $A_{LA} = (0.9 \pm 0.1)$~$\mu$m$^2$ and $A_{RA} = (0.68 \pm 0.08)$~$\mu$m$^2$ \footnote{The area of the two arms is obtained from the SEM image shown in Fig.~\ref{fig:1}(a). We have neglected a possible penetration of the magnetic field into the superconductor, because the London penetration depth of Nb is with 39 nm negligibly small \cite{Maxfield1965}.}. This suggests that the slowly varying envelope contribution $\Delta B_{slow}$ is exclusively caused by interference due to the finite extension of the two arms (Fraunhofer pattern).

Similarly, measuring the spacing between the supercurrent minima due to the fast oscillation, one obtains the periodicity $\Delta B_{fast} = (0.33 \pm 0.04)$~mT and the related area as $A_{fast} = \Phi_0 / \Delta B_{fast} = (6.27 \pm 0.76)$~$\mu$m$^2$. This is well above both the hole area ($A_{hole} = (1.7 \pm 0.2)$~$\mu$m$^2$) and even the whole mesa area ($A_{mesa} = (3.2 \pm 0.2)$~$\mu$m$^2$), indicating the presence of a magnetic focusing due to the Meissner effect, i.e.~the superconducting leads focus the magnetic field onto the mesa. It is this field piercing the central hole of the device which is responsible for the fast oscillations observed in the experiment. Given that focusing was not observed in the envelope (Fraunhofer) pattern, it must mainly affect the central hole of the mesa. This is reasonable, because the hole is well embedded between the leads, while both arms of the device are more external, and therefore less affected by the Meissner effect. We therefore define a focusing factor $F$ which is exactly 1 far from the leads, but in first approximation we set it to 1 also for the two arms of the device, i.e.~$F_{arm} = B_{arm} / B_{ext} = 1$, with $B_{ext}$ the externally applied magnetic field and $B_{arm}$ the effective magnetic field in the arm regions, while $F$ takes a value larger than unity in the hole region, $F_{hole} = B_{hole} / B_{ext} > 1$, with $B_{hole}$ the effective magnetic field in the hole region.

Figures~\ref{fig:2}(b) and (c) were measured in the same way as Fig.~\ref{fig:2}(a) but in the presence of negative voltages applied to the side--gates. With increasing negative gate voltage, the critical supercurrent intensity is suppressed and the device resistance increases. This is consistent with the fact that the side--gates increasingly deplete the 2DEG, reducing the width of the conducting channels from the outside in. The most evident effect is, however, the influence of the gate voltage on the slowly varying envelope. The amplitude of this component decreases with increasing negative gate voltage and essentially disappears for $-22.5$~V. This well agrees with the interpretation that, upon narrowing of the two channels, the behavior of the device becomes more and more SQUID--like --- in fact, an ideal SQUID exhibits no slowly varying envelope, but only a fast component.

For the data of Fig.~\ref{fig:2} we find
\begin{align*}
\Delta B_{fast}^{0 \text{~V}} &= (0.33 \pm 0.04) \text{~mT,} \\
\Delta B_{fast}^{-12.5 \text{~V}} &= (0.35 \pm 0.06) \text{~mT,} \\
\Delta B_{fast}^{-22.5 \text{~V}} &= (0.36 \pm 0.05) \text{~mT,}
\end{align*}
i.e.~the periodicity of the fast varying modulations is within error bars independent of gate voltage. This confirms that fast oscillations are exclusively caused by the magnetic flux piercing the central hole of the device, because otherwise their period should be affected by the reduction of the magnetic flux due to the shrinking of the area of the arms of the device imposed by gating. Thus we obtain $F_{hole} = A_{fast} / A_{hole} = 3.69 \pm 0.62$. We note that similar values of focusing factor have been recently reported in other hybrid structures \cite{Suominen2017,Monteiro2017}.

So far all measurements presented were performed in a symmetric configuration, with the same voltage applied to both side--gates. The device geometry, however, allows for an independent control of the voltage of each gate. These measurements are shown in Fig.~\ref{fig:3}. Interestingly, as we will show below, this allows to tune independently the supercurrent in each arm of the device, and also the effective width of each arm.

In the conditions of Fig.~\ref{fig:3}(a) and (c), the 2DEG in the left or right arm of the device, respectively, is nearly pinched off, while the other arm is open. Figs.~\ref{fig:3}(a,c) show that, although physically narrower, the right arm of the device can carry a higher supercurrent ($\approx 60$~nA for the central peak, approximately 2/3 of the initial total supercurrent), much larger than the left arm. Several factors might be responsible for this: different transparencies of the interfaces, or a higher initial carrier concentration in the right arm, or even the presence of impurities in the left arm, which eventually would locally reduce mobility and mean free path there.

\begin{figure*}[t]
   \includegraphics[width=\textwidth]{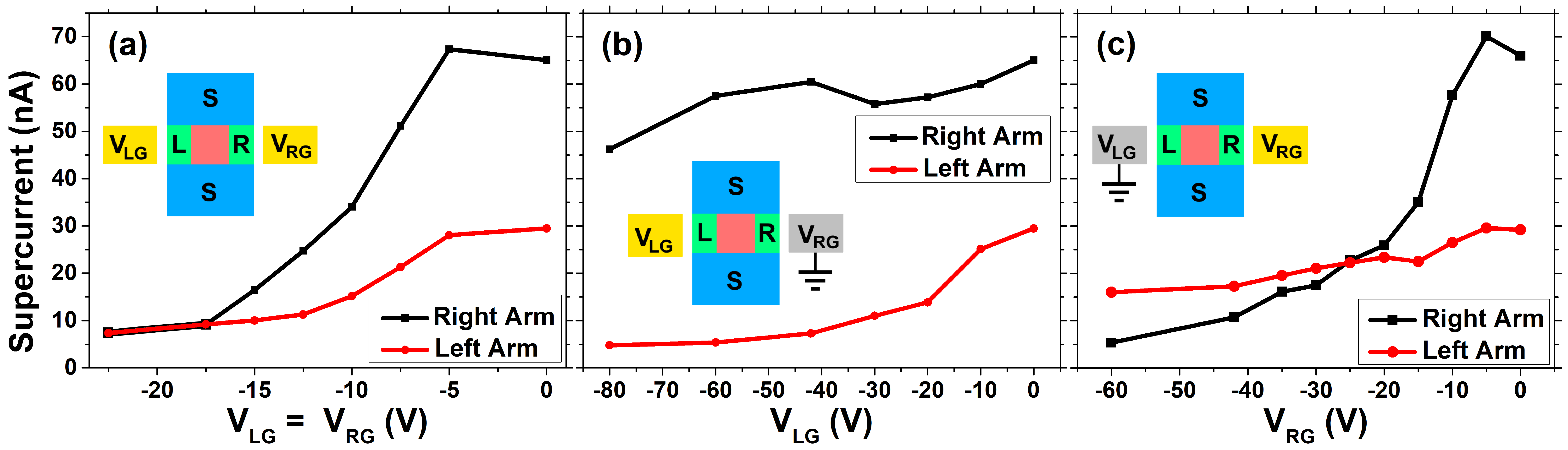}
   \caption{\label{fig:4} (a) Contribution to the total supercurrent from left arm (red) and right arm (black) when the same voltage is applied to both gates ($V_{LG} = V_{RG}$). The inset shows the schematics of the device with the same color scheme as in Fig.~\ref{fig:1}(a). (b,c) Same as before, but with asymmetric gating: (b) LG biased, RG grounded, (c) LG grounded, RG biased. $T = 315$~mK.}
\end{figure*}

Further inspection of Fig.~\ref{fig:3}(a) shows that the amplitude of the fast component  is much reduced as compared to Fig.~\ref{fig:2}(a), indicating that interference between the two arms of the device is strongly suppressed. The envelope modulation in Fig.~\ref{fig:3}(a) is more pronounced, and strongly resembles a Fraunhofer pattern with five visible lobes (one central lobe at zero magnetic field and four lateral lobes with reduced supercurrent). An analysis of the data in Fig.~\ref{fig:3}(a) yields the periodicity of the envelope modulation $\Delta B_{slow}^{-60 \text{~V}} = (2.98 \pm 0.44)$~mT and the related characteristic area $A_{slow}^{-60 \text{~V}} = (0.7 \pm 0.1)$~$\mu$m$^2$ which equals the area of the right arm $A_{RA} = (0.68 \pm 0.08)$~$\mu$m$^2$. This confirms that the envelope modulation is caused by interference effects due to the finite width of the right arm, the dominant one since the left one is almost pinched off. 

In the conditions of Figs.~\ref{fig:3}(a) and (c), the fast oscillations are reduced to a minimum amplitude, while the slow modulation is maximized. However, exploiting the large tunability of the device via the side--gates, it is also possible to set the opposite situation, as shown in Fig.~\ref{fig:3}(b). The striking feature here is the fact that in the central region of the pattern, the fast oscillations have a minimum value of zero supercurrent, i.e.~they extend all the way down to zero. As discussed below in detail, this implies that the contribution of the two arms to the supercurrent is perfectly balanced. This was obtained for an asymmetric bias configuration on the gates ($V_{LG} = 0$~V and $V_{RG} = -25$~V), consistently with the aforementioned asymmetry of the device. The three panels of Fig.~\ref{fig:3} thus illustrates the full tuneability of the device, from (a) a supercurrent flowing in the right arm, through (b) a situation where the two arms sustain an equal supercurrent, to (c) the supercurrent flowing in the left arm.

From the above discussion, the contribution of each device arm to the supercurrent as a function of bias applied to the side--gates can be extracted. An ideal SQUID with asymmetric and narrow arms has a supercurrent 
\begin{equation}
   \label{OurRealSQUID}
   I_c (\Phi) = \sqrt{\left(I_{c}^{RA} - I_{c}^{LA}\right)^2 + 4 I_{c}^{RA} I_{c}^{LA} \cos^2\left(\pi \Phi / \Phi_{0}\right)  },
\end{equation}
with $I_{c}^{RA}$ and $I_{c}^{LA}$ the critical current in the right and left arm, respectively, and $\Phi$ is the magnetic flux through the loop.

Since here the two arms of the device are extended JJs, also the contribution of the interference within each arm must be considered. The related envelope contribution is maximum for $\Phi = 0$ and slowly changes for increasing $\Phi$, while the SQUID contribution rapidly oscillates ($\Delta B_{slow} \gg \Delta B_{fast}$). Therefore, close to $\Phi = 0$, the envelope can be approximated as constant, and using Eq.~(\ref{OurRealSQUID}) we obtain
\begin{equation}
\begin{aligned}
  I_c (0) &= I_{c}^{RA} + I_{c}^{LA} \\
  I_c \left( \Phi_{0} / 2 \right) &= \left| I_{c}^{RA} - I_{c}^{LA} \right|
\end{aligned}
\end{equation}
Upon inversion, we get
\begin{equation}
\begin{aligned}
  I_{c}^{RA} &= \frac{I_c (0) + I_c \left( \Phi_{0} / 2 \right)}{2} \\
  I_{c}^{LA} &= \frac{I_c (0) - I_c \left( \Phi_{0} / 2 \right)}{2}
\end{aligned}
\end{equation}
for $I_c^{RA} > I_c^{LA}$. Therefore, from the minima and maxima of the fast oscillations close to zero, it is possible to calculate the contribution of each device arm to the total supercurrent.

Figure~\ref{fig:4}(a) shows the contribution of each device arm to the supercurrent for symmetrical gating ($V_{LG} = V_{RG}$). We note a progressive reduction in supercurrent in both arms with increasingly negative gate voltages, indicating that gating reduces the arm width. Going to more negative gate voltages, the device is therefore tuned from a non--ideal SQUID-like behavior with extended arms (see Fig.~\ref{fig:2}(a)), for which the interference pattern displays both a fast and a slow component, to a more ideal SQUID (see Fig.~\ref{fig:2}(c)), where the envelope has nearly disappeared. 

Figure~\ref{fig:4}(b) shows the supercurrent in the two arms as a function of a bias applied to LG, while RG is grounded. We note a strong reduction of supercurrent in the left arm (red curve) with increasing negative bias of LG, indicating a progressive reduction in arm width. Also the right arm is slightly affected by gating, especially for large negative bias. The progressive pinch--off of the left arm leads to the observation of a Fraunhofer pattern (see Fig.~\ref{fig:3}(a)). 

Finally, Fig.~\ref{fig:4}(c) shows the supercurrent in the two arms as a function of the bias applied to RG, while LG is grounded. Similarly to the previous case, a strong reduction of supercurrent is observed in the right arm close to the biased gate, with a detectable impact also on the other arm. The most relevant observation here is, however, the transition from an initial situation in which the right arm is dominant (corresponding to Fig.~\ref{fig:2}(a)), via a situation of balanced arms (Fig.~\ref{fig:3}(b)), to a situation of imbalance again, in which the left arm is now dominant (see Fig.~\ref{fig:3}(c)).

\section{\label{sec:Discussion}Discussion}

Experimental data can be analyzed using the following model of an extended Josephson junction. Assuming a rectangular junction, the supercurrent $I_c$ can be calculated as \cite{Tinkham1996}
\begin{equation}
\label{eq:Ic(B)}
I_c(B)=\max_{\Delta \phi}\left\lbrack \int_{-W/2}^{+W/2} J_{c}(x) \sin\gamma (x) dx \right\rbrack ,
\end{equation}
where $J_{c}(x)$ is the local critical-current density per unit length, $x$ the direction along the junction (perpendicular to the current flow), $\Delta \phi$ the phase difference between the superconductive leads, and $\gamma(x)$ the gauge-invariant local phase difference:
\begin{equation}
\gamma (x) = \Delta \phi - \frac{2\pi}{\Phi_0} \int_{}^{} \vec{A}(x,y)\cdot\vec{dl} ,
\end{equation}
with $y$ the direction along the current flow, $\vec{A}$ the vector potential, and the integration is performed from one electrode of the weak link to the other. The measured value of interface transparency ($\tau = 0.50$) allows to use a sine for the current-phase relation (CPR) in Eq.~(\ref{eq:Ic(B)}), since when the junction transparency decreases from 1, the CPR rapidly goes towards a sine function \cite{Schapers2001}.

Since $L \ll W$ we can consider only trajectories perpendicular to the interface \cite{Schapers2001}, and therefore we obtain:
\begin{equation}
\gamma (x) = \Delta \phi - \frac{2\pi}{\Phi_0} \int_{0}^{L} A_y(x,y)dy .
\end{equation}
Under the reasonable assumption that, in this geometry, $A_y$ is constant along the $y$ direction, we get 
\begin{equation}
\gamma (x) = \Delta \phi - \frac{2 \pi}{\Phi_0} A_y(x) L,
\end{equation}
where $A_y(x)$ is linked to the magnetic field by:
\begin{equation}
\frac{\partial A_y (x)}{\partial x} =B_{ext}F(x),
\end{equation}
with $F(x)$ the focusing factor profile.

\begin{figure}[t!]
   \includegraphics[width=0.8\columnwidth]{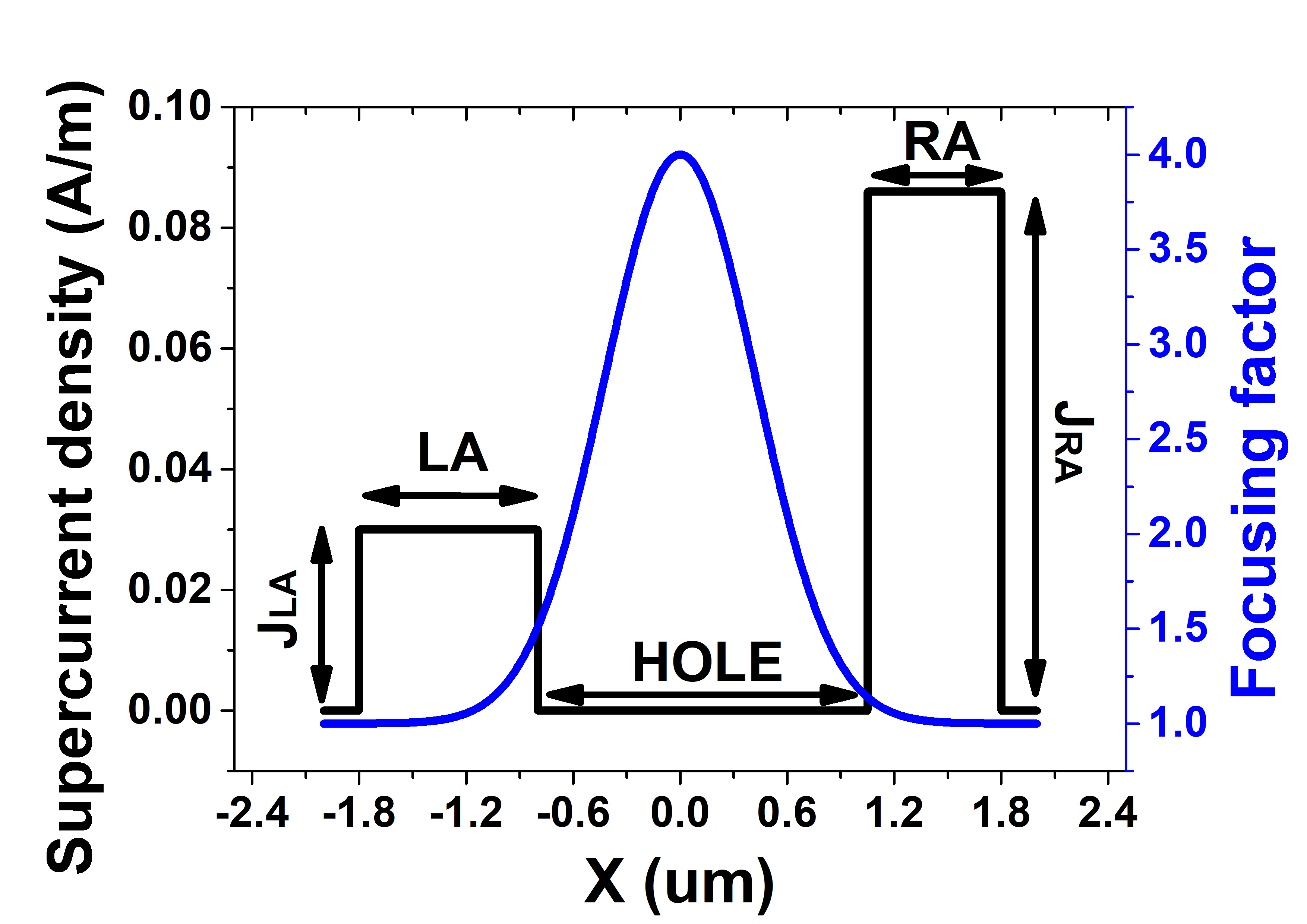}
   \caption{\label{fig:5}Supercurrent density (black) and focusing factor (blue) as a function of position. $LA$ e $RA$ are the width of the left and right interferometer arm, respectively, while $J_{LA}$ and $J_{RA}$ are the corresponding supercurrent densities. The focusing factor profile was used for all simulations presented in Fig.~\ref{fig:6}, while the supercurrent density profile is the one used for $V_{LG} = V_{RG}= 0$~V [Fig.~\ref{fig:6}(a)].}
\end{figure}

In order to compare simulations based on this model with the experimental data, a few assumptions were made. For $J_{c}(x)$ we chose $J_{c}(x) = 0$ in the hole area (or, more generally, outside the area of the two interferometer arms), and a constant value within the two arms (with current density $J_{LA}$ and $J_{RA}$ in the left and right arm, respectively). This approximation neglects any inhomogeneity in the arms or at their interfaces and allows, for $V_{LG} = V_{RG}= 0$~V, to determine the width of the two arms from the SEM measurement (Fig.~\ref{fig:1}). This is further justified since surface depletion effects in InAs (at the edge of the mesa) can be safely neglected \cite{Noguchi1991,Olsson1996}. We therefore obtain:
\begin{equation}
J_{c}(x)=\left\{
\begin{array}{ccl}
J_{LA} & ~ & -1.8 <x<-0.8~\mu{\rm m} \\
0  & ~ & -0.8 <x<1.05~\mu{\rm m} \\
J_{RA} & ~ & 1.05 <x<1.8~\mu{\rm m} 
\end{array}.
\right.
\end{equation}
On the other hand, the value of the current density in the two arms is known, see Fig.~\ref{fig:4}. Here, for $V_{LG} = V_{RG}= 0$~V, $J_{LA} = (0.029 \pm 0.001)$~A/m and $J_{RA} = (0.086 \pm 0.002)$~A/m, calculated from the measured values of $I_{LA} = (29.5 \pm 1.0)$~nA and $I_{RA} = (65.0 \pm 1.0)$~nA. Figure~\ref{fig:5} shows in black the resulting current density profile. We stress that there are no free parameters; all values are obtained from measurement. Instead, for the analysis of the measurements using the side--gates, the effective widths of the two arms is unknown and therefore treated as fitting parameters. 

\begin{figure*}[tbh]
   \includegraphics[width=\textwidth]{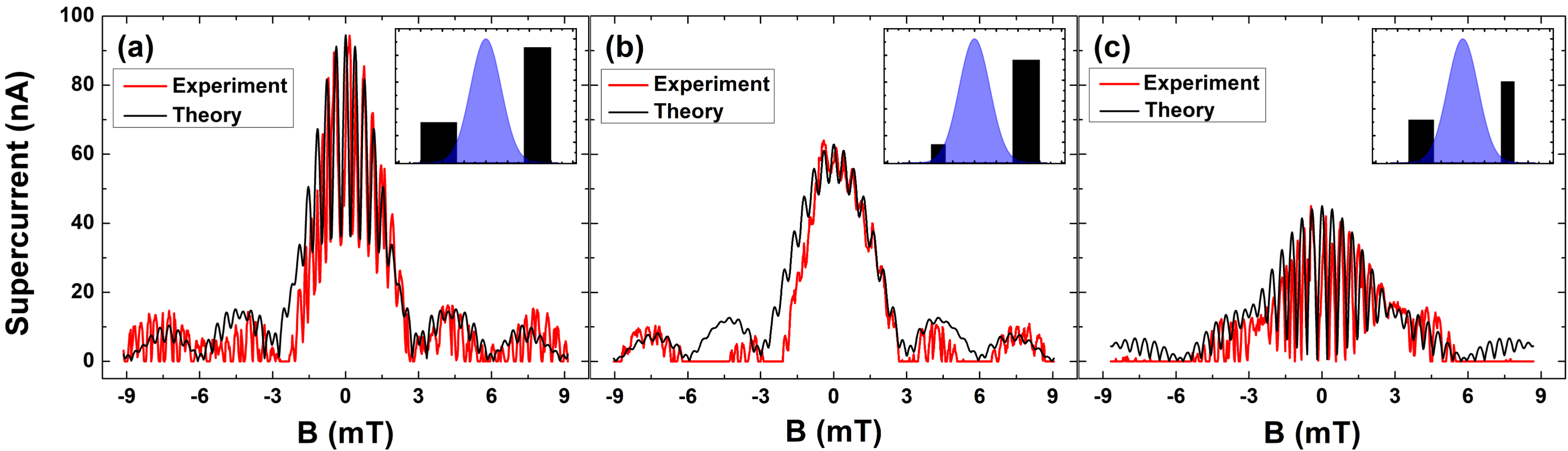}
   \caption{\label{fig:6}Supercurrent as a function of the applied external magnetic field $B$. Experimental data are plotted in red, while the results of the simulations are plotted in black. (a) corresponds to the measurement shown in Fig.~\ref{fig:2}(a) [$V_{LG} = V_{RG}= 0$~V], (b) to Fig.~\ref{fig:3}(a) [$V_{LG} = -60$~V and $V_{RG}= 0$~V], and (c) to Fig.~\ref{fig:3}(b) [$V_{LG} = 0$V and $V_{RG}= -25$~V].}
\end{figure*}

The precise form of the focusing factor, larger than one in the hole region, is not so critical, because the supercurrent in the hole region is zero. Here, it is important that the integral under the focusing factor function equals the total flux that pierces the hole area. The selected function should also decay rapidly so as to allow for a fast transition from the hole area ($F > 1$) to the arms ($F = 1$). We have therefore described the focusing factor profile by a Gaussian function, using the following expression:
\begin{equation}
F(x) = 1 + (F_{max}-1)e^{-\left(x / F_{\sigma} \right) ^2},
\end{equation}
with $F_{max}$ the amplitude and $F_{\sigma}$ the width of the Gaussian. The values of these two parameters (which are not independent, since the integral under the Gaussian is known) were determined by fitting the data of Fig.~\ref{fig:2}(a) taken at $V_{LG} = V_{RG}= 0$~V and then used for all other values of gate voltage. Their values are $F_{max} = 4.0 \pm 0.4$ and $F_{\sigma} = (600 \pm 35)$~nm, consistently with  the estimate discussed in section~\ref{sec:ExperimentalResults} ($F_{hole} = 3.8 \pm 0.9$) and with the requirement that the Gaussian should be confined to the hole area. Figure~\ref{fig:5} shows in blue the resulting focusing factor profile.

The experimental data of Fig.~\ref{fig:2}(a) are shown together with the simulation in Fig.~\ref{fig:6}(a). The agreement between simulation and experiment is good. The shape and amplitude of the envelope are well reproduced, and even more strikingly, the fast oscillations in the central lobe are quantitatively captured. In particular, while for the data we find $\Delta B_{fast}^{data} = (0.33 \pm 0.04)$~~mT and $\Delta B_{slow}^{data} = (2.89 \pm 0.25)$~mT, in the simulation we obtain $\Delta B_{fast}^{sim} = 0.34$~mT and $\Delta B_{slow}^{sim} = 2.98$~mT.

Moreover, Fig.~\ref{fig:6}(b) shows a simulation of the data from Fig.~\ref{fig:3}(a), while Fig.~\ref{fig:6}(c) shows a simulation of the data from Fig.~\ref{fig:3}(b). Also for these two cases, a good agreement between simulation and experiment is found. From the simulations, the effective width of each arm, and thus the supercurrent density, as a function of applied gate voltage is obtained (we remind that the latter are not free parameters, since the total supercurrent for each arm is known). The results of the simulations are summarized in Table~\ref{tbl1}. These results consistently demonstrate that the side--gates reduce the effective width of the device arms and the corresponding supercurrent density.

\begin{table*}[bth]
  \caption{\label{tbl1} Parameters of the simulations shown in Fig.~\ref{fig:6}.}
  \begin{ruledtabular}
  \begin{tabular}{l|cccccc}
	                    & $V_{LG}$ (V) & $V_{RG}$ (V) & $LA$ ($\mu$m)   & $RA$ ($\mu$m)   & $J_{LA}$ (A/m)    & $J_{RA}$ (A/m)    \\ \hline
	Fig.~\ref{fig:6}(a) & 0            & 0	          & 1.0             & 0.75            & $0.029 \pm 0.001$ & $0.086 \pm 0.002$ \\
	Fig.~\ref{fig:6}(b) & $-60$        & 0            & $0.40 \pm 0.05$ & $0.75 \pm 0.01$ & $0.013 \pm 0.004$ & $0.076 \pm 0.002$ \\
	Fig.~\ref{fig:6}(c) & 0            & $-25$        & $0.70 \pm 0.05$ & $0.38 \pm 0.04$ & $0.032 \pm 0.004$ & $0.059 \pm 0.009$ \\
  \end{tabular}
  \end{ruledtabular}
\end{table*}

Thus, the model captures all relevant features of the device. Furthermore, it yields information on the device that would be difficult to obtain otherwise, e.g. the effective width of the device arms. This may be used for supercurrent spectroscopy or a spatial mapping of the supercurrent density in each arm.

\section{\label{sec:Summary}Summary}

We have demonstrated a fully-tunable hybrid semiconductor/superconductor device, in which the width, area, and supercurrent of the two arms of an extended trenched Josephson junction can be independently controlled with high precision. With no electrostatic gating, the magnetic interference pattern exhibits fast SQUID--like oscillations on top of a slow Fraunhofer--like envelope. The fast oscillations are contributed by the interference between two arms of the device, while the slow envelope is due to the finite width of each arm. In particular, the contribution of the two device arms to the supercurrent can be perfectly balanced by suitably biasing side--gates. This allows to tune the device characteristics from a SQUID--like behaviour, with narrow arms, to that of an extended single--arm Josephson junction showing a pure Fraunnhofer--pattern. Interestingly, the transition between these limits is obtained in a continuous manner on the same device via electrostatic gating, without the need for additional external in-plane magnetic fields. The wide tunability offered by this new geometry, and its electrostatic control, is very promising for applications, and moreover it constitutes an easily scalable platform.

\section{\label{sec:Methods}Methods}

The InAs quantum well based heterostructure (sample number HM3586) was grown by means of molecular beam epitaxy on a GaAs (001) substrate on top of which a series of 50-nm-thick In$_{1-x}$Al$_{x}$As layers was deposited (the concentration in Al varies from $x = 0.85$ in the first layer to $x = 0.25$ in the last one). A 4-nm-thick InAs QW is then interposed between two 5.5-nm-thick In$_{0.75}$Ga$_{0.25}$As layers and asymmetric In$_{0.75}$Al$_{0.25}$As barriers, with the lower one delta-doped with Si at a 13~nm setback \cite{Capotondi2005,Amado2014}. The center of the InAs well is placed at 57.5~nm from the top surface (see Fig.~\ref{fig:1}(b)). The sheet electron density $n \simeq 3.74 \times 10^{11}$~cm$^{-2}$, the mobility $\mu \simeq 2 \times 10^5$~cm$^2$/(Vs), and the effective electron mass $m \simeq 0.03 m_{e}$ were extracted from low-temperature Shubnikov-de-Haas oscillations measurements (at 250~mK). We can, therefore, estimate the mean free path of the 2DEG as $\ell \simeq 2$~$\mu$m \cite{Amado2014}.

The fabrication of the SNS device required a sequence of mutually aligned steps of electron beam lithography (EBL), see Refs.~\cite{Amado2013,Fornieri2013a,Fornieri2013,Guiducci2014,Amado2014} for further details. With the first lithography we defined the mesa region of the 2DEG, i.e.~the rectangular central island of the device that acts as the N region. To this end, a negative resist bilayer was spin coated on the surface of the sample and served as the mask defining the 2DEG-region. The uncovered part of the heterostructure was then removed ($\approx 350$~nm deep) by dipping in a H$_2$O:H$_2$SO$_4$:H$_2$O$_2$ solution (chemical wet etching). The superconducting parts of the device were designed by the second step of EBL. Prior to the sputter deposition of a 150-nm-thick Nb film, the surface was cleaned from undesired oxide layer with a dip in a HF:H$_2$O solution and a low-energy Ar$^+$ milling in the sputtering chamber. Differently from previous works \cite{Amado2013,Guiducci2014,Amado2014} we added two metallic side--gates insulated from the heterostructure by a 100-nm-thick hydrogen silsesquioxane (HSQ) layer (third lithography) followed by the thermal evaporation of a Ti/Au bilayer (5/145~nm) in a fourth lithographic step. Side--gates patterned this way allow for a dramatic suppression of the leakage towards the heterostructure even at high gate voltage. Finally, we used the tungsten tip of a commercial tuning-fork based AFM (Attocube AFM III) to scratch the central region of the mesa. In this way, the 2DEG in the scratched region was removed.

Transport measurements in magnetic field are performed by fixing the voltage applied to each side--gate to the values reported in the various plots. No positive gate voltages were used in the experiment. To collect the data, we fixed the magnetic field, and then for each value of magnetic field, we took an $I-V$ curve using the same setup as depicted in Fig.~\ref{fig:1}(a). To plot the differential resistance shown in Fig.~\ref{fig:2}(a), we took the numerical derivative of each $I-V$ curve. $I-V$ curves were typically taken with magnetic field steps of $0.025$~mT. In all measurements, both the magnetic field and the source-drain current are swept from positive to negative values. As a result of the chosen current sweep direction, the retrapping (switching) current is always at positive (negative) values of the current. The more pronounced hysteretic behavior observed in Fig.~\ref{fig:2}(a) and Fig.~\ref{fig:3}(a) is consistent with the larger supercurrent observed in these measurements.

\begin{acknowledgments}
We are indebted to Francesco Giazotto for his support in the initial phase of this research. We thank Alessandro Braggio, Clemens Winkelmann, Lucia Sorba, and Rosario Fazio for useful discussions. S.~G.~acknowledges support by Fondazione Silvio Tronchetti Provera. M.~C.~acknowledges support from the Quant-Eranet project "SuperTop". S.~H.~acknowledges support from Scuola Normale Superiore, project SNS16\_B\_HEUN---004155. We acknowledge support from SNS-WIS joint lab 'QUANTRA'. Financial support from the CNR in the framework of the agreement on scientific collaboration between CNR and CNRS (France) and CNR and CONICET (Argentina) is acknowledged. Furthermore, this work was financially supported by EC through the ERC Advanced Grant No.~670173.
\end{acknowledgments}

\bibliography{SQUID-lib}

\begin{thebibliography}{58}%
\makeatletter
\providecommand \@ifxundefined [1]{%
 \@ifx{#1\undefined}
}%
\providecommand \@ifnum [1]{%
 \ifnum #1\expandafter \@firstoftwo
 \else \expandafter \@secondoftwo
 \fi
}%
\providecommand \@ifx [1]{%
 \ifx #1\expandafter \@firstoftwo
 \else \expandafter \@secondoftwo
 \fi
}%
\providecommand \natexlab [1]{#1}%
\providecommand \enquote  [1]{``#1''}%
\providecommand \bibnamefont  [1]{#1}%
\providecommand \bibfnamefont [1]{#1}%
\providecommand \citenamefont [1]{#1}%
\providecommand \href@noop [0]{\@secondoftwo}%
\providecommand \href [0]{\begingroup \@sanitize@url \@href}%
\providecommand \@href[1]{\@@startlink{#1}\@@href}%
\providecommand \@@href[1]{\endgroup#1\@@endlink}%
\providecommand \@sanitize@url [0]{\catcode `\\12\catcode `\$12\catcode
  `\&12\catcode `\#12\catcode `\^12\catcode `\_12\catcode `\%12\relax}%
\providecommand \@@startlink[1]{}%
\providecommand \@@endlink[0]{}%
\providecommand \url  [0]{\begingroup\@sanitize@url \@url }%
\providecommand \@url [1]{\endgroup\@href {#1}{\urlprefix }}%
\providecommand \urlprefix  [0]{URL }%
\providecommand \Eprint [0]{\href }%
\providecommand \doibase [0]{http://dx.doi.org/}%
\providecommand \selectlanguage [0]{\@gobble}%
\providecommand \bibinfo  [0]{\@secondoftwo}%
\providecommand \bibfield  [0]{\@secondoftwo}%
\providecommand \translation [1]{[#1]}%
\providecommand \BibitemOpen [0]{}%
\providecommand \bibitemStop [0]{}%
\providecommand \bibitemNoStop [0]{.\EOS\space}%
\providecommand \EOS [0]{\spacefactor3000\relax}%
\providecommand \BibitemShut  [1]{\csname bibitem#1\endcsname}%
\let\auto@bib@innerbib\@empty
\bibitem [{\citenamefont {Barone}\ and\ \citenamefont
  {Patern\`o}(1982)}]{Barone1982}%
  \BibitemOpen
  \bibfield  {author} {\bibinfo {author} {\bibfnamefont {A.}~\bibnamefont
  {Barone}}\ and\ \bibinfo {author} {\bibfnamefont {G.}~\bibnamefont
  {Patern\`o}},\ }\href@noop {} {\emph {\bibinfo {title} {{Physics and
  Applications of the Josephson Effect}}}}\ (\bibinfo  {publisher} {John Wiley
  \& Sons, Inc.},\ \bibinfo {year} {1982})\BibitemShut {NoStop}%
\bibitem [{\citenamefont {Tinkham}(1996)}]{Tinkham1996}%
  \BibitemOpen
  \bibfield  {author} {\bibinfo {author} {\bibfnamefont {M.}~\bibnamefont
  {Tinkham}},\ }\href@noop {} {\emph {\bibinfo {title} {{Introduction to
  Superconductivity}}}}\ (\bibinfo  {publisher} {McGraw-Hill},\ \bibinfo {year}
  {1996})\BibitemShut {NoStop}%
\bibitem [{\citenamefont {Ronzani}\ \emph {et~al.}(2014)\citenamefont
  {Ronzani}, \citenamefont {Altimiras},\ and\ \citenamefont
  {Giazotto}}]{Ronzani2014}%
  \BibitemOpen
  \bibfield  {author} {\bibinfo {author} {\bibfnamefont {A.}~\bibnamefont
  {Ronzani}}, \bibinfo {author} {\bibfnamefont {C.}~\bibnamefont {Altimiras}},
  \ and\ \bibinfo {author} {\bibfnamefont {F.}~\bibnamefont {Giazotto}},\
  }\href@noop {} {\bibfield  {journal} {\bibinfo  {journal} {Phys. Rev.
  Applied}\ }\textbf {\bibinfo {volume} {2}},\ \bibinfo {pages} {024005}
  (\bibinfo {year} {2014})}\BibitemShut {NoStop}%
\bibitem [{\citenamefont {Jabdaraghi}\ \emph {et~al.}(2018)\citenamefont
  {Jabdaraghi}, \citenamefont {Peltonen}, \citenamefont {Golubev},\ and\
  \citenamefont {Pekola}}]{Jabdaraghi2018}%
  \BibitemOpen
  \bibfield  {author} {\bibinfo {author} {\bibfnamefont {R.~N.}\ \bibnamefont
  {Jabdaraghi}}, \bibinfo {author} {\bibfnamefont {J.~T.}\ \bibnamefont
  {Peltonen}}, \bibinfo {author} {\bibfnamefont {D.~S.}\ \bibnamefont
  {Golubev}}, \ and\ \bibinfo {author} {\bibfnamefont {J.~P.}\ \bibnamefont
  {Pekola}},\ }\href@noop {} {\bibfield  {journal} {\bibinfo  {journal} {J. Low
  Temp. Phys.}\ }\textbf {\bibinfo {volume} {191}},\ \bibinfo {pages} {344}
  (\bibinfo {year} {2018})}\BibitemShut {NoStop}%
\bibitem [{\citenamefont {Govenius}\ \emph {et~al.}(2016)\citenamefont
  {Govenius}, \citenamefont {Lake}, \citenamefont {Tan},\ and\ \citenamefont
  {M\"ott\"onen}}]{Govenius2016}%
  \BibitemOpen
  \bibfield  {author} {\bibinfo {author} {\bibfnamefont {J.}~\bibnamefont
  {Govenius}}, \bibinfo {author} {\bibfnamefont {R.~E.}\ \bibnamefont {Lake}},
  \bibinfo {author} {\bibfnamefont {K.~Y.}\ \bibnamefont {Tan}}, \ and\
  \bibinfo {author} {\bibfnamefont {M.}~\bibnamefont {M\"ott\"onen}},\
  }\href@noop {} {\bibfield  {journal} {\bibinfo  {journal} {Phys. Rev. Lett.}\
  }\textbf {\bibinfo {volume} {117}},\ \bibinfo {pages} {030802} (\bibinfo
  {year} {2016})}\BibitemShut {NoStop}%
\bibitem [{\citenamefont {Wendin}(2017)}]{Wendin2017}%
  \BibitemOpen
  \bibfield  {author} {\bibinfo {author} {\bibfnamefont {G.}~\bibnamefont
  {Wendin}},\ }\href@noop {} {\bibfield  {journal} {\bibinfo  {journal} {Rep.
  Prog. Phys.}\ }\textbf {\bibinfo {volume} {80}},\ \bibinfo {pages} {106001}
  (\bibinfo {year} {2017})}\BibitemShut {NoStop}%
\bibitem [{\citenamefont {Mourik}\ \emph {et~al.}(2012)\citenamefont {Mourik},
  \citenamefont {Zuo}, \citenamefont {Frolov}, \citenamefont {Plissard},
  \citenamefont {Bakkers},\ and\ \citenamefont {Kouwenhoven}}]{Mourik2012}%
  \BibitemOpen
  \bibfield  {author} {\bibinfo {author} {\bibfnamefont {V.}~\bibnamefont
  {Mourik}}, \bibinfo {author} {\bibfnamefont {K.}~\bibnamefont {Zuo}},
  \bibinfo {author} {\bibfnamefont {S.~M.}\ \bibnamefont {Frolov}}, \bibinfo
  {author} {\bibfnamefont {S.~R.}\ \bibnamefont {Plissard}}, \bibinfo {author}
  {\bibfnamefont {E.~P. A.~M.}\ \bibnamefont {Bakkers}}, \ and\ \bibinfo
  {author} {\bibfnamefont {L.~P.}\ \bibnamefont {Kouwenhoven}},\ }\href@noop {}
  {\bibfield  {journal} {\bibinfo  {journal} {{Science}}\ }\textbf {\bibinfo
  {volume} {336}},\ \bibinfo {pages} {1003} (\bibinfo {year}
  {2012})}\BibitemShut {NoStop}%
\bibitem [{\citenamefont {Beenakker}(2013)}]{Beenakker2013}%
  \BibitemOpen
  \bibfield  {author} {\bibinfo {author} {\bibfnamefont {C.~W.~J.}\
  \bibnamefont {Beenakker}},\ }\href@noop {} {\bibfield  {journal} {\bibinfo
  {journal} {Annual Review of Condensed Mat\-ter Physics}\ }\textbf {\bibinfo
  {volume} {4}},\ \bibinfo {pages} {113} (\bibinfo {year} {2013})}\BibitemShut
  {NoStop}%
\bibitem [{\citenamefont {Mong}\ \emph {et~al.}(2014)\citenamefont {Mong},
  \citenamefont {Clarke}, \citenamefont {Alicea}, \citenamefont {Lindner},
  \citenamefont {Fendley}, \citenamefont {Nayak}, \citenamefont {Oreg},
  \citenamefont {Stern}, \citenamefont {Berg}, \citenamefont {Shtengel},\ and\
  \citenamefont {Fisher}}]{Mong2014}%
  \BibitemOpen
  \bibfield  {author} {\bibinfo {author} {\bibfnamefont {R.~S.~K.}\
  \bibnamefont {Mong}}, \bibinfo {author} {\bibfnamefont {D.~J.}\ \bibnamefont
  {Clarke}}, \bibinfo {author} {\bibfnamefont {J.}~\bibnamefont {Alicea}},
  \bibinfo {author} {\bibfnamefont {N.~H.}\ \bibnamefont {Lindner}}, \bibinfo
  {author} {\bibfnamefont {P.}~\bibnamefont {Fendley}}, \bibinfo {author}
  {\bibfnamefont {C.}~\bibnamefont {Nayak}}, \bibinfo {author} {\bibfnamefont
  {Y.}~\bibnamefont {Oreg}}, \bibinfo {author} {\bibfnamefont {A.}~\bibnamefont
  {Stern}}, \bibinfo {author} {\bibfnamefont {E.}~\bibnamefont {Berg}},
  \bibinfo {author} {\bibfnamefont {K.}~\bibnamefont {Shtengel}}, \ and\
  \bibinfo {author} {\bibfnamefont {M.~P.~A.}\ \bibnamefont {Fisher}},\
  }\href@noop {} {\bibfield  {journal} {\bibinfo  {journal} {Phys. Rev. X}\
  }\textbf {\bibinfo {volume} {4}},\ \bibinfo {pages} {011036} (\bibinfo {year}
  {2014})}\BibitemShut {NoStop}%
\bibitem [{\citenamefont {Albrecht}\ \emph {et~al.}(2016)\citenamefont
  {Albrecht}, \citenamefont {Higginbotham}, \citenamefont {Madsen},
  \citenamefont {Kuemmeth}, \citenamefont {Jespersen}, \citenamefont {Nygard},
  \citenamefont {Krogstrup},\ and\ \citenamefont {Marcus}}]{Albrecht2016}%
  \BibitemOpen
  \bibfield  {author} {\bibinfo {author} {\bibfnamefont {S.~M.}\ \bibnamefont
  {Albrecht}}, \bibinfo {author} {\bibfnamefont {A.~P.}\ \bibnamefont
  {Higginbotham}}, \bibinfo {author} {\bibfnamefont {M.}~\bibnamefont
  {Madsen}}, \bibinfo {author} {\bibfnamefont {F.}~\bibnamefont {Kuemmeth}},
  \bibinfo {author} {\bibfnamefont {T.~S.}\ \bibnamefont {Jespersen}}, \bibinfo
  {author} {\bibfnamefont {J.}~\bibnamefont {Nygard}}, \bibinfo {author}
  {\bibfnamefont {P.}~\bibnamefont {Krogstrup}}, \ and\ \bibinfo {author}
  {\bibfnamefont {C.~M.}\ \bibnamefont {Marcus}},\ }\href@noop {} {\bibfield
  {journal} {\bibinfo  {journal} {{Nature}}\ }\textbf {\bibinfo {volume}
  {531}},\ \bibinfo {pages} {206} (\bibinfo {year} {2016})}\BibitemShut
  {NoStop}%
\bibitem [{\citenamefont {Ishikawa}\ and\ \citenamefont
  {Fukuyama}(1999)}]{Ishikawa1999}%
  \BibitemOpen
  \bibfield  {author} {\bibinfo {author} {\bibfnamefont {Y.}~\bibnamefont
  {Ishikawa}}\ and\ \bibinfo {author} {\bibfnamefont {H.}~\bibnamefont
  {Fukuyama}},\ }\href@noop {} {\bibfield  {journal} {\bibinfo  {journal} {J.
  Phys. Soc. Jpn.}\ }\textbf {\bibinfo {volume} {68}},\ \bibinfo {pages} {954}
  (\bibinfo {year} {1999})}\BibitemShut {NoStop}%
\bibitem [{\citenamefont {Silaev}(2017)}]{Silaev2017}%
  \BibitemOpen
  \bibfield  {author} {\bibinfo {author} {\bibfnamefont {M.~A.}\ \bibnamefont
  {Silaev}},\ }\href@noop {} {\bibfield  {journal} {\bibinfo  {journal}
  {arXiv:1708.07467 [cond-mat.supr-con]}\ } (\bibinfo {year}
  {2017})}\BibitemShut {NoStop}%
\bibitem [{\citenamefont {Nakamura}\ \emph {et~al.}(1999)\citenamefont
  {Nakamura}, \citenamefont {Pashkin},\ and\ \citenamefont
  {Tsai}}]{Nakamura1999}%
  \BibitemOpen
  \bibfield  {author} {\bibinfo {author} {\bibfnamefont {Y.}~\bibnamefont
  {Nakamura}}, \bibinfo {author} {\bibfnamefont {Y.~A.}\ \bibnamefont
  {Pashkin}}, \ and\ \bibinfo {author} {\bibfnamefont {J.~S.}\ \bibnamefont
  {Tsai}},\ }\href@noop {} {\bibfield  {journal} {\bibinfo  {journal}
  {{Nature}}\ }\textbf {\bibinfo {volume} {398}},\ \bibinfo {pages} {786}
  (\bibinfo {year} {1999})}\BibitemShut {NoStop}%
\bibitem [{\citenamefont {Friedman}\ \emph {et~al.}(2000)\citenamefont
  {Friedman}, \citenamefont {Patel}, \citenamefont {Chen}, \citenamefont
  {Tolpygo},\ and\ \citenamefont {Lukens}}]{Friedman2000}%
  \BibitemOpen
  \bibfield  {author} {\bibinfo {author} {\bibfnamefont {J.~R.}\ \bibnamefont
  {Friedman}}, \bibinfo {author} {\bibfnamefont {V.}~\bibnamefont {Patel}},
  \bibinfo {author} {\bibfnamefont {W.}~\bibnamefont {Chen}}, \bibinfo {author}
  {\bibfnamefont {S.~K.}\ \bibnamefont {Tolpygo}}, \ and\ \bibinfo {author}
  {\bibfnamefont {J.~E.}\ \bibnamefont {Lukens}},\ }\href@noop {} {\bibfield
  {journal} {\bibinfo  {journal} {{Nature}}\ }\textbf {\bibinfo {volume}
  {406}},\ \bibinfo {pages} {43} (\bibinfo {year} {2000})}\BibitemShut
  {NoStop}%
\bibitem [{\citenamefont {Yu}\ \emph {et~al.}(2002)\citenamefont {Yu},
  \citenamefont {Han}, \citenamefont {Chu}, \citenamefont {Chu},\ and\
  \citenamefont {Wang}}]{Yu2002}%
  \BibitemOpen
  \bibfield  {author} {\bibinfo {author} {\bibfnamefont {Y.}~\bibnamefont
  {Yu}}, \bibinfo {author} {\bibfnamefont {S.}~\bibnamefont {Han}}, \bibinfo
  {author} {\bibfnamefont {X.}~\bibnamefont {Chu}}, \bibinfo {author}
  {\bibfnamefont {S.-I.}\ \bibnamefont {Chu}}, \ and\ \bibinfo {author}
  {\bibfnamefont {Z.}~\bibnamefont {Wang}},\ }\href@noop {} {\bibfield
  {journal} {\bibinfo  {journal} {{Science}}\ }\textbf {\bibinfo {volume}
  {296}},\ \bibinfo {pages} {889} (\bibinfo {year} {2002})}\BibitemShut
  {NoStop}%
\bibitem [{\citenamefont {Belzig}\ \emph {et~al.}(1999)\citenamefont {Belzig},
  \citenamefont {Wilhelm}, \citenamefont {Bruder}, \citenamefont {Sch\"on},\
  and\ \citenamefont {Zaikin}}]{Belzig1999}%
  \BibitemOpen
  \bibfield  {author} {\bibinfo {author} {\bibfnamefont {W.}~\bibnamefont
  {Belzig}}, \bibinfo {author} {\bibfnamefont {F.~K.}\ \bibnamefont {Wilhelm}},
  \bibinfo {author} {\bibfnamefont {C.}~\bibnamefont {Bruder}}, \bibinfo
  {author} {\bibfnamefont {G.}~\bibnamefont {Sch\"on}}, \ and\ \bibinfo
  {author} {\bibfnamefont {A.~D.}\ \bibnamefont {Zaikin}},\ }\href@noop {}
  {\bibfield  {journal} {\bibinfo  {journal} {Superlattices Microstruct.}\
  }\textbf {\bibinfo {volume} {25}},\ \bibinfo {pages} {1251} (\bibinfo {year}
  {1999})}\BibitemShut {NoStop}%
\bibitem [{\citenamefont {Courtois}\ \emph {et~al.}(1999)\citenamefont
  {Courtois}, \citenamefont {Gandit}, \citenamefont {Pannetier},\ and\
  \citenamefont {Mailly}}]{Courtois1999a}%
  \BibitemOpen
  \bibfield  {author} {\bibinfo {author} {\bibfnamefont {H.}~\bibnamefont
  {Courtois}}, \bibinfo {author} {\bibfnamefont {P.}~\bibnamefont {Gandit}},
  \bibinfo {author} {\bibfnamefont {B.}~\bibnamefont {Pannetier}}, \ and\
  \bibinfo {author} {\bibfnamefont {D.}~\bibnamefont {Mailly}},\ }\href
  {\doibase https://doi.org/10.1006/spmi.1999.0711} {\bibfield  {journal}
  {\bibinfo  {journal} {Superlattices Microstruct.}\ }\textbf {\bibinfo
  {volume} {25}},\ \bibinfo {pages} {721} (\bibinfo {year} {1999})}\BibitemShut
  {NoStop}%
\bibitem [{\citenamefont {Amado}\ \emph {et~al.}(2014)\citenamefont {Amado},
  \citenamefont {Fornieri}, \citenamefont {Biasiol}, \citenamefont {Sorba},\
  and\ \citenamefont {Giazotto}}]{Amado2014}%
  \BibitemOpen
  \bibfield  {author} {\bibinfo {author} {\bibfnamefont {M.}~\bibnamefont
  {Amado}}, \bibinfo {author} {\bibfnamefont {A.}~\bibnamefont {Fornieri}},
  \bibinfo {author} {\bibfnamefont {G.}~\bibnamefont {Biasiol}}, \bibinfo
  {author} {\bibfnamefont {L.}~\bibnamefont {Sorba}}, \ and\ \bibinfo {author}
  {\bibfnamefont {F.}~\bibnamefont {Giazotto}},\ }\href@noop {} {\bibfield
  {journal} {\bibinfo  {journal} {Appl. Phys. Lett.}\ }\textbf {\bibinfo
  {volume} {104}},\ \bibinfo {pages} {242604} (\bibinfo {year}
  {2014})}\BibitemShut {NoStop}%
\bibitem [{\citenamefont {Fornieri}\ \emph {et~al.}(2013)\citenamefont
  {Fornieri}, \citenamefont {Amado}, \citenamefont {Carillo}, \citenamefont
  {Dolcini}, \citenamefont {Biasiol}, \citenamefont {Sorba}, \citenamefont
  {Pellegrini},\ and\ \citenamefont {Giazotto}}]{Fornieri2013}%
  \BibitemOpen
  \bibfield  {author} {\bibinfo {author} {\bibfnamefont {A.}~\bibnamefont
  {Fornieri}}, \bibinfo {author} {\bibfnamefont {M.}~\bibnamefont {Amado}},
  \bibinfo {author} {\bibfnamefont {F.}~\bibnamefont {Carillo}}, \bibinfo
  {author} {\bibfnamefont {F.}~\bibnamefont {Dolcini}}, \bibinfo {author}
  {\bibfnamefont {G.}~\bibnamefont {Biasiol}}, \bibinfo {author} {\bibfnamefont
  {L.}~\bibnamefont {Sorba}}, \bibinfo {author} {\bibfnamefont
  {V.}~\bibnamefont {Pellegrini}}, \ and\ \bibinfo {author} {\bibfnamefont
  {F.}~\bibnamefont {Giazotto}},\ }\href@noop {} {\bibfield  {journal}
  {\bibinfo  {journal} {Nanotechnology}\ }\textbf {\bibinfo {volume} {24}},\
  \bibinfo {pages} {245201} (\bibinfo {year} {2013})}\BibitemShut {NoStop}%
\bibitem [{\citenamefont {Drachmann}\ \emph {et~al.}(2017)\citenamefont
  {Drachmann}, \citenamefont {Suominen}, \citenamefont {Kjaergaard},
  \citenamefont {Shojaei}, \citenamefont {Palmstr\o{}m}, \citenamefont
  {Marcus},\ and\ \citenamefont {Nichele}}]{Drachmann2017}%
  \BibitemOpen
  \bibfield  {author} {\bibinfo {author} {\bibfnamefont {A.~C.~C.}\
  \bibnamefont {Drachmann}}, \bibinfo {author} {\bibfnamefont {H.~J.}\
  \bibnamefont {Suominen}}, \bibinfo {author} {\bibfnamefont {M.}~\bibnamefont
  {Kjaergaard}}, \bibinfo {author} {\bibfnamefont {B.}~\bibnamefont {Shojaei}},
  \bibinfo {author} {\bibfnamefont {C.~J.}\ \bibnamefont {Palmstr\o{}m}},
  \bibinfo {author} {\bibfnamefont {C.~M.}\ \bibnamefont {Marcus}}, \ and\
  \bibinfo {author} {\bibfnamefont {F.}~\bibnamefont {Nichele}},\ }\href@noop
  {} {\bibfield  {journal} {\bibinfo  {journal} {Nano Lett.}\ }\textbf
  {\bibinfo {volume} {17}},\ \bibinfo {pages} {1200} (\bibinfo {year}
  {2017})}\BibitemShut {NoStop}%
\bibitem [{\citenamefont {Takayanagi}\ \emph {et~al.}(1995)\citenamefont
  {Takayanagi}, \citenamefont {Akazaki},\ and\ \citenamefont
  {Nitta}}]{Takayanagi1995}%
  \BibitemOpen
  \bibfield  {author} {\bibinfo {author} {\bibfnamefont {H.}~\bibnamefont
  {Takayanagi}}, \bibinfo {author} {\bibfnamefont {T.}~\bibnamefont {Akazaki}},
  \ and\ \bibinfo {author} {\bibfnamefont {J.}~\bibnamefont {Nitta}},\ }\href
  {\doibase 10.1103/PhysRevLett.75.3533} {\bibfield  {journal} {\bibinfo
  {journal} {Phys. Rev. Lett.}\ }\textbf {\bibinfo {volume} {75}},\ \bibinfo
  {pages} {3533} (\bibinfo {year} {1995})}\BibitemShut {NoStop}%
\bibitem [{\citenamefont {Shabani}\ \emph {et~al.}(2016)\citenamefont
  {Shabani}, \citenamefont {Kjaergaard}, \citenamefont {Suominen},
  \citenamefont {Kim}, \citenamefont {Nichele}, \citenamefont {Pakrouski},
  \citenamefont {Stankevic}, \citenamefont {Lutchyn}, \citenamefont
  {Krogstrup}, \citenamefont {Feidenhans'l}, \citenamefont {Kraemer},
  \citenamefont {Nayak}, \citenamefont {Troyer}, \citenamefont {Marcus},\ and\
  \citenamefont {Palmstr\o{}m}}]{Shabani2016}%
  \BibitemOpen
  \bibfield  {author} {\bibinfo {author} {\bibfnamefont {J.}~\bibnamefont
  {Shabani}}, \bibinfo {author} {\bibfnamefont {M.}~\bibnamefont {Kjaergaard}},
  \bibinfo {author} {\bibfnamefont {H.~J.}\ \bibnamefont {Suominen}}, \bibinfo
  {author} {\bibfnamefont {Y.}~\bibnamefont {Kim}}, \bibinfo {author}
  {\bibfnamefont {F.}~\bibnamefont {Nichele}}, \bibinfo {author} {\bibfnamefont
  {K.}~\bibnamefont {Pakrouski}}, \bibinfo {author} {\bibfnamefont
  {T.}~\bibnamefont {Stankevic}}, \bibinfo {author} {\bibfnamefont {R.~M.}\
  \bibnamefont {Lutchyn}}, \bibinfo {author} {\bibfnamefont {P.}~\bibnamefont
  {Krogstrup}}, \bibinfo {author} {\bibfnamefont {R.}~\bibnamefont
  {Feidenhans'l}}, \bibinfo {author} {\bibfnamefont {S.}~\bibnamefont
  {Kraemer}}, \bibinfo {author} {\bibfnamefont {C.}~\bibnamefont {Nayak}},
  \bibinfo {author} {\bibfnamefont {M.}~\bibnamefont {Troyer}}, \bibinfo
  {author} {\bibfnamefont {C.~M.}\ \bibnamefont {Marcus}}, \ and\ \bibinfo
  {author} {\bibfnamefont {C.~J.}\ \bibnamefont {Palmstr\o{}m}},\ }\href@noop
  {} {\bibfield  {journal} {\bibinfo  {journal} {Phys. Rev. B}\ }\textbf
  {\bibinfo {volume} {93}},\ \bibinfo {pages} {155402} (\bibinfo {year}
  {2016})}\BibitemShut {NoStop}%
\bibitem [{\citenamefont {Kjaergaard}\ \emph {et~al.}(2016)\citenamefont
  {Kjaergaard}, \citenamefont {Nichele}, \citenamefont {Suominen},
  \citenamefont {Nowak}, \citenamefont {Wimmer}, \citenamefont {Akhmerov},
  \citenamefont {Folk}, \citenamefont {Flensberg}, \citenamefont {Shabani},
  \citenamefont {Palmstr\o{}m},\ and\ \citenamefont {Marcus}}]{Kjaergaard2016}%
  \BibitemOpen
  \bibfield  {author} {\bibinfo {author} {\bibfnamefont {M.}~\bibnamefont
  {Kjaergaard}}, \bibinfo {author} {\bibfnamefont {F.}~\bibnamefont {Nichele}},
  \bibinfo {author} {\bibfnamefont {H.~J.}\ \bibnamefont {Suominen}}, \bibinfo
  {author} {\bibfnamefont {M.~P.}\ \bibnamefont {Nowak}}, \bibinfo {author}
  {\bibfnamefont {M.}~\bibnamefont {Wimmer}}, \bibinfo {author} {\bibfnamefont
  {A.~R.}\ \bibnamefont {Akhmerov}}, \bibinfo {author} {\bibfnamefont {J.~A.}\
  \bibnamefont {Folk}}, \bibinfo {author} {\bibfnamefont {K.}~\bibnamefont
  {Flensberg}}, \bibinfo {author} {\bibfnamefont {J.}~\bibnamefont {Shabani}},
  \bibinfo {author} {\bibfnamefont {C.~J.}\ \bibnamefont {Palmstr\o{}m}}, \
  and\ \bibinfo {author} {\bibfnamefont {C.~M.}\ \bibnamefont {Marcus}},\
  }\href@noop {} {\bibfield  {journal} {\bibinfo  {journal} {Nat. Commun.}\
  }\textbf {\bibinfo {volume} {7}},\ \bibinfo {pages} {12841} (\bibinfo {year}
  {2016})}\BibitemShut {NoStop}%
\bibitem [{\citenamefont {Kjaergaard}\ \emph {et~al.}(2017)\citenamefont
  {Kjaergaard}, \citenamefont {Suominen}, \citenamefont {Nowak}, \citenamefont
  {Akhmerov}, \citenamefont {Shabani}, \citenamefont {Palmstr\o{}m},
  \citenamefont {Nichele},\ and\ \citenamefont {Marcus}}]{Kjaergaard2017}%
  \BibitemOpen
  \bibfield  {author} {\bibinfo {author} {\bibfnamefont {M.}~\bibnamefont
  {Kjaergaard}}, \bibinfo {author} {\bibfnamefont {H.~J.}\ \bibnamefont
  {Suominen}}, \bibinfo {author} {\bibfnamefont {M.~P.}\ \bibnamefont {Nowak}},
  \bibinfo {author} {\bibfnamefont {A.~R.}\ \bibnamefont {Akhmerov}}, \bibinfo
  {author} {\bibfnamefont {J.}~\bibnamefont {Shabani}}, \bibinfo {author}
  {\bibfnamefont {C.~J.}\ \bibnamefont {Palmstr\o{}m}}, \bibinfo {author}
  {\bibfnamefont {F.}~\bibnamefont {Nichele}}, \ and\ \bibinfo {author}
  {\bibfnamefont {C.~M.}\ \bibnamefont {Marcus}},\ }\href@noop {} {\bibfield
  {journal} {\bibinfo  {journal} {Phys. Rev. Applied}\ }\textbf {\bibinfo
  {volume} {7}},\ \bibinfo {pages} {034029} (\bibinfo {year}
  {2017})}\BibitemShut {NoStop}%
\bibitem [{\citenamefont {Goffman}\ \emph {et~al.}(2017)\citenamefont
  {Goffman}, \citenamefont {Urbina}, \citenamefont {Pothier}, \citenamefont
  {Nyg\aa{}rd}, \citenamefont {Marcus},\ and\ \citenamefont
  {Krogstrup}}]{Goffman2017}%
  \BibitemOpen
  \bibfield  {author} {\bibinfo {author} {\bibfnamefont {M.~F.}\ \bibnamefont
  {Goffman}}, \bibinfo {author} {\bibfnamefont {C.}~\bibnamefont {Urbina}},
  \bibinfo {author} {\bibfnamefont {H.}~\bibnamefont {Pothier}}, \bibinfo
  {author} {\bibfnamefont {J.}~\bibnamefont {Nyg\aa{}rd}}, \bibinfo {author}
  {\bibfnamefont {C.~M.}\ \bibnamefont {Marcus}}, \ and\ \bibinfo {author}
  {\bibfnamefont {P.}~\bibnamefont {Krogstrup}},\ }\href@noop {} {\bibfield
  {journal} {\bibinfo  {journal} {New J. Phys.}\ }\textbf {\bibinfo {volume}
  {19}},\ \bibinfo {pages} {092002} (\bibinfo {year} {2017})}\BibitemShut
  {NoStop}%
\bibitem [{\citenamefont {Sestoft}\ \emph {et~al.}(2018)\citenamefont
  {Sestoft}, \citenamefont {Kanne}, \citenamefont {Gejl}, \citenamefont {von
  Soosten}, \citenamefont {Yodh}, \citenamefont {Sherman}, \citenamefont
  {Tarasinski}, \citenamefont {Wimmer}, \citenamefont {Johnson}, \citenamefont
  {Deng}, \citenamefont {Nyg\aa{}rd}, \citenamefont {Jespersen}, \citenamefont
  {Marcus},\ and\ \citenamefont {Krogstrup}}]{Sestoft2018}%
  \BibitemOpen
  \bibfield  {author} {\bibinfo {author} {\bibfnamefont {J.~E.}\ \bibnamefont
  {Sestoft}}, \bibinfo {author} {\bibfnamefont {T.}~\bibnamefont {Kanne}},
  \bibinfo {author} {\bibfnamefont {A.~N.}\ \bibnamefont {Gejl}}, \bibinfo
  {author} {\bibfnamefont {M.}~\bibnamefont {von Soosten}}, \bibinfo {author}
  {\bibfnamefont {J.~S.}\ \bibnamefont {Yodh}}, \bibinfo {author}
  {\bibfnamefont {D.}~\bibnamefont {Sherman}}, \bibinfo {author} {\bibfnamefont
  {B.}~\bibnamefont {Tarasinski}}, \bibinfo {author} {\bibfnamefont
  {M.}~\bibnamefont {Wimmer}}, \bibinfo {author} {\bibfnamefont
  {E.}~\bibnamefont {Johnson}}, \bibinfo {author} {\bibfnamefont
  {M.}~\bibnamefont {Deng}}, \bibinfo {author} {\bibfnamefont {J.}~\bibnamefont
  {Nyg\aa{}rd}}, \bibinfo {author} {\bibfnamefont {T.~S.}\ \bibnamefont
  {Jespersen}}, \bibinfo {author} {\bibfnamefont {C.~M.}\ \bibnamefont
  {Marcus}}, \ and\ \bibinfo {author} {\bibfnamefont {P.}~\bibnamefont
  {Krogstrup}},\ }\href@noop {} {\bibfield  {journal} {\bibinfo  {journal}
  {Phys. Rev. Materials}\ }\textbf {\bibinfo {volume} {2}},\ \bibinfo {pages}
  {044202} (\bibinfo {year} {2018})}\BibitemShut {NoStop}%
\bibitem [{\citenamefont {Casparis}\ \emph {et~al.}(2017)\citenamefont
  {Casparis}, \citenamefont {Connolly}, \citenamefont {Kjaergaard},
  \citenamefont {Pearson}, \citenamefont {Kringh\o{}j}, \citenamefont {Larsen},
  \citenamefont {Kuemmeth}, \citenamefont {Wang}, \citenamefont {Thomas},
  \citenamefont {Gronin}, \citenamefont {Gardner}, \citenamefont {Manfra},
  \citenamefont {Marcus},\ and\ \citenamefont {Petersson}}]{Casparis2017}%
  \BibitemOpen
  \bibfield  {author} {\bibinfo {author} {\bibfnamefont {L.}~\bibnamefont
  {Casparis}}, \bibinfo {author} {\bibfnamefont {M.~R.}\ \bibnamefont
  {Connolly}}, \bibinfo {author} {\bibfnamefont {M.}~\bibnamefont
  {Kjaergaard}}, \bibinfo {author} {\bibfnamefont {N.~J.}\ \bibnamefont
  {Pearson}}, \bibinfo {author} {\bibfnamefont {A.}~\bibnamefont
  {Kringh\o{}j}}, \bibinfo {author} {\bibfnamefont {T.~W.}\ \bibnamefont
  {Larsen}}, \bibinfo {author} {\bibfnamefont {F.}~\bibnamefont {Kuemmeth}},
  \bibinfo {author} {\bibfnamefont {T.}~\bibnamefont {Wang}}, \bibinfo {author}
  {\bibfnamefont {C.}~\bibnamefont {Thomas}}, \bibinfo {author} {\bibfnamefont
  {S.}~\bibnamefont {Gronin}}, \bibinfo {author} {\bibfnamefont {G.~C.}\
  \bibnamefont {Gardner}}, \bibinfo {author} {\bibfnamefont {M.~J.}\
  \bibnamefont {Manfra}}, \bibinfo {author} {\bibfnamefont {C.~M.}\
  \bibnamefont {Marcus}}, \ and\ \bibinfo {author} {\bibfnamefont {K.~D.}\
  \bibnamefont {Petersson}},\ }\href@noop {} {\bibfield  {journal} {\bibinfo
  {journal} {Nat. Nanotechnol.}\ }\textbf {\bibinfo {volume} {13}},\ \bibinfo
  {pages} {915} (\bibinfo {year} {2017})}\BibitemShut {NoStop}%
\bibitem [{\citenamefont {Suominen}\ \emph {et~al.}(2017)\citenamefont
  {Suominen}, \citenamefont {Danon}, \citenamefont {Kjaergaard}, \citenamefont
  {Flensberg}, \citenamefont {Shabani}, \citenamefont {Palmstr\o{}m},
  \citenamefont {Nichele},\ and\ \citenamefont {Marcus}}]{Suominen2017}%
  \BibitemOpen
  \bibfield  {author} {\bibinfo {author} {\bibfnamefont {H.~J.}\ \bibnamefont
  {Suominen}}, \bibinfo {author} {\bibfnamefont {J.}~\bibnamefont {Danon}},
  \bibinfo {author} {\bibfnamefont {M.}~\bibnamefont {Kjaergaard}}, \bibinfo
  {author} {\bibfnamefont {K.}~\bibnamefont {Flensberg}}, \bibinfo {author}
  {\bibfnamefont {J.}~\bibnamefont {Shabani}}, \bibinfo {author} {\bibfnamefont
  {C.~J.}\ \bibnamefont {Palmstr\o{}m}}, \bibinfo {author} {\bibfnamefont
  {F.}~\bibnamefont {Nichele}}, \ and\ \bibinfo {author} {\bibfnamefont
  {C.~M.}\ \bibnamefont {Marcus}},\ }\href@noop {} {\bibfield  {journal}
  {\bibinfo  {journal} {Phys. Rev. B}\ }\textbf {\bibinfo {volume} {95}},\
  \bibinfo {pages} {035307} (\bibinfo {year} {2017})}\BibitemShut {NoStop}%
\bibitem [{\citenamefont {Amado}\ \emph {et~al.}(2013)\citenamefont {Amado},
  \citenamefont {Fornieri}, \citenamefont {Carillo}, \citenamefont {Biasiol},
  \citenamefont {Sorba}, \citenamefont {Pellegrini},\ and\ \citenamefont
  {Giazotto}}]{Amado2013}%
  \BibitemOpen
  \bibfield  {author} {\bibinfo {author} {\bibfnamefont {M.}~\bibnamefont
  {Amado}}, \bibinfo {author} {\bibfnamefont {A.}~\bibnamefont {Fornieri}},
  \bibinfo {author} {\bibfnamefont {F.}~\bibnamefont {Carillo}}, \bibinfo
  {author} {\bibfnamefont {G.}~\bibnamefont {Biasiol}}, \bibinfo {author}
  {\bibfnamefont {L.}~\bibnamefont {Sorba}}, \bibinfo {author} {\bibfnamefont
  {V.}~\bibnamefont {Pellegrini}}, \ and\ \bibinfo {author} {\bibfnamefont
  {F.}~\bibnamefont {Giazotto}},\ }\href@noop {} {\bibfield  {journal}
  {\bibinfo  {journal} {Phys. Rev. B}\ }\textbf {\bibinfo {volume} {87}},\
  \bibinfo {pages} {134506} (\bibinfo {year} {2013})}\BibitemShut {NoStop}%
\bibitem [{\citenamefont {Paajaste}\ \emph {et~al.}(2015)\citenamefont
  {Paajaste}, \citenamefont {Amado}, \citenamefont {Roddaro}, \citenamefont
  {Bergeret}, \citenamefont {Ercolani}, \citenamefont {Sorba},\ and\
  \citenamefont {Giazotto}}]{Paajaste2015}%
  \BibitemOpen
  \bibfield  {author} {\bibinfo {author} {\bibfnamefont {J.}~\bibnamefont
  {Paajaste}}, \bibinfo {author} {\bibfnamefont {M.}~\bibnamefont {Amado}},
  \bibinfo {author} {\bibfnamefont {S.}~\bibnamefont {Roddaro}}, \bibinfo
  {author} {\bibfnamefont {F.~S.}\ \bibnamefont {Bergeret}}, \bibinfo {author}
  {\bibfnamefont {D.}~\bibnamefont {Ercolani}}, \bibinfo {author}
  {\bibfnamefont {L.}~\bibnamefont {Sorba}}, \ and\ \bibinfo {author}
  {\bibfnamefont {F.}~\bibnamefont {Giazotto}},\ }\href@noop {} {\bibfield
  {journal} {\bibinfo  {journal} {Nano Lett.}\ }\textbf {\bibinfo {volume}
  {15}},\ \bibinfo {pages} {1803} (\bibinfo {year} {2015})}\BibitemShut
  {NoStop}%
\bibitem [{\citenamefont {Guiducci}\ \emph {et~al.}(2019)\citenamefont
  {Guiducci}, \citenamefont {Carrega}, \citenamefont {Biasiol}, \citenamefont
  {Sorba}, \citenamefont {Beltram},\ and\ \citenamefont {Heun}}]{Guiducci2018}%
  \BibitemOpen
  \bibfield  {author} {\bibinfo {author} {\bibfnamefont {S.}~\bibnamefont
  {Guiducci}}, \bibinfo {author} {\bibfnamefont {M.}~\bibnamefont {Carrega}},
  \bibinfo {author} {\bibfnamefont {G.}~\bibnamefont {Biasiol}}, \bibinfo
  {author} {\bibfnamefont {L.}~\bibnamefont {Sorba}}, \bibinfo {author}
  {\bibfnamefont {F.}~\bibnamefont {Beltram}}, \ and\ \bibinfo {author}
  {\bibfnamefont {S.}~\bibnamefont {Heun}},\ }\href@noop {} {\bibfield
  {journal} {\bibinfo  {journal} {{Phys. Status Solidi RRL}}\ }\textbf
  {\bibinfo {volume} {13}},\ \bibinfo {pages} {1800222} (\bibinfo {year}
  {2019})}\BibitemShut {NoStop}%
\bibitem [{\citenamefont {Ke}\ \emph {et~al.}(2019)\citenamefont {Ke},
  \citenamefont {Moehle}, \citenamefont {de~Vries}, \citenamefont {Thomas},
  \citenamefont {Metti}, \citenamefont {Guinn}, \citenamefont {Kallaher},
  \citenamefont {Lodari}, \citenamefont {Scappucci}, \citenamefont {Wang},
  \citenamefont {Diaz}, \citenamefont {Gardner}, \citenamefont {Manfra},\ and\
  \citenamefont {Goswami}}]{Ke2019}%
  \BibitemOpen
  \bibfield  {author} {\bibinfo {author} {\bibfnamefont {C.~T.}\ \bibnamefont
  {Ke}}, \bibinfo {author} {\bibfnamefont {C.~M.}\ \bibnamefont {Moehle}},
  \bibinfo {author} {\bibfnamefont {F.~K.}\ \bibnamefont {de~Vries}}, \bibinfo
  {author} {\bibfnamefont {C.}~\bibnamefont {Thomas}}, \bibinfo {author}
  {\bibfnamefont {S.}~\bibnamefont {Metti}}, \bibinfo {author} {\bibfnamefont
  {C.~R.}\ \bibnamefont {Guinn}}, \bibinfo {author} {\bibfnamefont
  {R.}~\bibnamefont {Kallaher}}, \bibinfo {author} {\bibfnamefont
  {M.}~\bibnamefont {Lodari}}, \bibinfo {author} {\bibfnamefont
  {G.}~\bibnamefont {Scappucci}}, \bibinfo {author} {\bibfnamefont
  {T.}~\bibnamefont {Wang}}, \bibinfo {author} {\bibfnamefont {R.~E.}\
  \bibnamefont {Diaz}}, \bibinfo {author} {\bibfnamefont {G.~C.}\ \bibnamefont
  {Gardner}}, \bibinfo {author} {\bibfnamefont {M.~J.}\ \bibnamefont {Manfra}},
  \ and\ \bibinfo {author} {\bibfnamefont {S.}~\bibnamefont {Goswami}},\
  }\href@noop {} {\bibfield  {journal} {\bibinfo  {journal} {arXiv:1902.10742
  [cond-mat.mes-hall]}\ } (\bibinfo {year} {2019})}\BibitemShut {NoStop}%
\bibitem [{\citenamefont {Seredinski}\ \emph {et~al.}(2019)\citenamefont
  {Seredinski}, \citenamefont {Draelos}, \citenamefont {Arnault}, \citenamefont
  {Wei}, \citenamefont {Li}, \citenamefont {Watanabe}, \citenamefont
  {Taniguchi}, \citenamefont {Amet},\ and\ \citenamefont
  {Finkelstein}}]{Seredinski2019}%
  \BibitemOpen
  \bibfield  {author} {\bibinfo {author} {\bibfnamefont {A.}~\bibnamefont
  {Seredinski}}, \bibinfo {author} {\bibfnamefont {A.~W.}\ \bibnamefont
  {Draelos}}, \bibinfo {author} {\bibfnamefont {E.~G.}\ \bibnamefont
  {Arnault}}, \bibinfo {author} {\bibfnamefont {M.-T.}\ \bibnamefont {Wei}},
  \bibinfo {author} {\bibfnamefont {H.}~\bibnamefont {Li}}, \bibinfo {author}
  {\bibfnamefont {K.}~\bibnamefont {Watanabe}}, \bibinfo {author}
  {\bibfnamefont {T.}~\bibnamefont {Taniguchi}}, \bibinfo {author}
  {\bibfnamefont {F.}~\bibnamefont {Amet}}, \ and\ \bibinfo {author}
  {\bibfnamefont {G.}~\bibnamefont {Finkelstein}},\ }\href@noop {} {\bibfield
  {journal} {\bibinfo  {journal} {arXiv:1901.05928 [cond-mat.mes-hall]}\ }
  (\bibinfo {year} {2019})}\BibitemShut {NoStop}%
\bibitem [{\citenamefont {Monteiro}\ \emph {et~al.}(2017)\citenamefont
  {Monteiro}, \citenamefont {Groenendijk}, \citenamefont {Manca}, \citenamefont
  {Mulazimoglu}, \citenamefont {Goswami}, \citenamefont {Blanter},
  \citenamefont {Vandersypen},\ and\ \citenamefont {Caviglia}}]{Monteiro2017}%
  \BibitemOpen
  \bibfield  {author} {\bibinfo {author} {\bibfnamefont {A.~M. R. V.~L.}\
  \bibnamefont {Monteiro}}, \bibinfo {author} {\bibfnamefont {D.~J.}\
  \bibnamefont {Groenendijk}}, \bibinfo {author} {\bibfnamefont
  {N.}~\bibnamefont {Manca}}, \bibinfo {author} {\bibfnamefont
  {E.}~\bibnamefont {Mulazimoglu}}, \bibinfo {author} {\bibfnamefont
  {S.}~\bibnamefont {Goswami}}, \bibinfo {author} {\bibfnamefont
  {Y.}~\bibnamefont {Blanter}}, \bibinfo {author} {\bibfnamefont {L.~M.~K.}\
  \bibnamefont {Vandersypen}}, \ and\ \bibinfo {author} {\bibfnamefont {A.~D.}\
  \bibnamefont {Caviglia}},\ }\href@noop {} {\bibfield  {journal} {\bibinfo
  {journal} {Nano Letters}\ }\textbf {\bibinfo {volume} {17}},\ \bibinfo
  {pages} {715} (\bibinfo {year} {2017})}\BibitemShut {NoStop}%
\bibitem [{\citenamefont {Thompson}\ \emph {et~al.}(2017)\citenamefont
  {Thompson}, \citenamefont {Ben~Shalom}, \citenamefont {Geim}, \citenamefont
  {Matthews}, \citenamefont {White}, \citenamefont {Melhem}, \citenamefont
  {Pashkin}, \citenamefont {Haley},\ and\ \citenamefont
  {Prance}}]{Thompson2017}%
  \BibitemOpen
  \bibfield  {author} {\bibinfo {author} {\bibfnamefont {M.~D.}\ \bibnamefont
  {Thompson}}, \bibinfo {author} {\bibfnamefont {M.}~\bibnamefont
  {Ben~Shalom}}, \bibinfo {author} {\bibfnamefont {A.~K.}\ \bibnamefont
  {Geim}}, \bibinfo {author} {\bibfnamefont {A.~J.}\ \bibnamefont {Matthews}},
  \bibinfo {author} {\bibfnamefont {J.}~\bibnamefont {White}}, \bibinfo
  {author} {\bibfnamefont {Z.}~\bibnamefont {Melhem}}, \bibinfo {author}
  {\bibfnamefont {Y.~A.}\ \bibnamefont {Pashkin}}, \bibinfo {author}
  {\bibfnamefont {R.~P.}\ \bibnamefont {Haley}}, \ and\ \bibinfo {author}
  {\bibfnamefont {J.~R.}\ \bibnamefont {Prance}},\ }\href@noop {} {\bibfield
  {journal} {\bibinfo  {journal} {Applied Physics Letters}\ }\textbf {\bibinfo
  {volume} {110}},\ \bibinfo {pages} {162602} (\bibinfo {year}
  {2017})}\BibitemShut {NoStop}%
\bibitem [{\citenamefont {Fornieri}(2013)}]{Fornieri2013a}%
  \BibitemOpen
  \bibfield  {author} {\bibinfo {author} {\bibfnamefont {A.}~\bibnamefont
  {Fornieri}},\ }\emph {\bibinfo {title} {Josephson effect in ballistic
  semiconductor nanostructures}},\ \href@noop {} {Master's thesis},\ \bibinfo
  {school} {University of Pisa} (\bibinfo {year} {2013})\BibitemShut {NoStop}%
\bibitem [{\citenamefont {Guiducci}(2014)}]{Guiducci2014}%
  \BibitemOpen
  \bibfield  {author} {\bibinfo {author} {\bibfnamefont {S.}~\bibnamefont
  {Guiducci}},\ }\emph {\bibinfo {title} {Electron transport and scanning gate
  microscopy studies on ballistic hybrid {SNS} junctions}},\ \href@noop {}
  {Master's thesis},\ \bibinfo  {school} {University of Pisa} (\bibinfo {year}
  {2014})\BibitemShut {NoStop}%
\bibitem [{\citenamefont {Ashcroft}\ and\ \citenamefont
  {Mermin}(1976)}]{Ashcroft1976}%
  \BibitemOpen
  \bibfield  {author} {\bibinfo {author} {\bibfnamefont {N.~W.}\ \bibnamefont
  {Ashcroft}}\ and\ \bibinfo {author} {\bibfnamefont {N.~D.}\ \bibnamefont
  {Mermin}},\ }\href@noop {} {\emph {\bibinfo {title} {{Solid State
  Physics}}}}\ (\bibinfo  {publisher} {Saunders College},\ \bibinfo {year}
  {1976})\BibitemShut {NoStop}%
\bibitem [{\citenamefont {Mur}\ \emph {et~al.}(1996)\citenamefont {Mur},
  \citenamefont {Harmans}, \citenamefont {Mooij}, \citenamefont {Carlin},
  \citenamefont {Rudra},\ and\ \citenamefont {Ilegems}}]{Mur1996}%
  \BibitemOpen
  \bibfield  {author} {\bibinfo {author} {\bibfnamefont {L.~C.}\ \bibnamefont
  {Mur}}, \bibinfo {author} {\bibfnamefont {C.~J. P.~M.}\ \bibnamefont
  {Harmans}}, \bibinfo {author} {\bibfnamefont {J.~E.}\ \bibnamefont {Mooij}},
  \bibinfo {author} {\bibfnamefont {J.~F.}\ \bibnamefont {Carlin}}, \bibinfo
  {author} {\bibfnamefont {A.}~\bibnamefont {Rudra}}, \ and\ \bibinfo {author}
  {\bibfnamefont {M.}~\bibnamefont {Ilegems}},\ }\href@noop {} {\bibfield
  {journal} {\bibinfo  {journal} {{Phys. Rev. B}}\ }\textbf {\bibinfo {volume}
  {54}},\ \bibinfo {pages} {R2327} (\bibinfo {year} {1996})}\BibitemShut
  {NoStop}%
\bibitem [{\citenamefont {Grosso}\ and\ \citenamefont
  {Parravicini}(2000)}]{Grosso2000}%
  \BibitemOpen
  \bibfield  {author} {\bibinfo {author} {\bibfnamefont {G.}~\bibnamefont
  {Grosso}}\ and\ \bibinfo {author} {\bibfnamefont {G.~P.}\ \bibnamefont
  {Parravicini}},\ }\href@noop {} {\emph {\bibinfo {title} {Solid state
  physics}}}\ (\bibinfo  {publisher} {Academic Press},\ \bibinfo {year}
  {2000})\BibitemShut {NoStop}%
\bibitem [{\citenamefont {Schapers}(2001)}]{Schapers2001}%
  \BibitemOpen
  \bibfield  {author} {\bibinfo {author} {\bibfnamefont {T.}~\bibnamefont
  {Schapers}},\ }\href@noop {} {\emph {\bibinfo {title} {Superconductor /
  semiconductor junctions}}}\ (\bibinfo  {publisher} {Sprin\-ger},\ \bibinfo
  {year} {2001})\BibitemShut {NoStop}%
\bibitem [{\citenamefont {Courtois}\ \emph {et~al.}(2008)\citenamefont
  {Courtois}, \citenamefont {Meschke}, \citenamefont {Peltonen},\ and\
  \citenamefont {Pekola}}]{Courtois2008}%
  \BibitemOpen
  \bibfield  {author} {\bibinfo {author} {\bibfnamefont {H.}~\bibnamefont
  {Courtois}}, \bibinfo {author} {\bibfnamefont {M.}~\bibnamefont {Meschke}},
  \bibinfo {author} {\bibfnamefont {J.~T.}\ \bibnamefont {Peltonen}}, \ and\
  \bibinfo {author} {\bibfnamefont {J.~P.}\ \bibnamefont {Pekola}},\ }\href
  {\doibase 10.1103/PhysRevLett.101.067002} {\bibfield  {journal} {\bibinfo
  {journal} {Phys. Rev. Lett.}\ }\textbf {\bibinfo {volume} {101}},\ \bibinfo
  {pages} {067002} (\bibinfo {year} {2008})}\BibitemShut {NoStop}%
\bibitem [{\citenamefont {Marsh}\ \emph {et~al.}(1994)\citenamefont {Marsh},
  \citenamefont {Williams},\ and\ \citenamefont {Ahmed}}]{Marsh1994}%
  \BibitemOpen
  \bibfield  {author} {\bibinfo {author} {\bibfnamefont {A.~M.}\ \bibnamefont
  {Marsh}}, \bibinfo {author} {\bibfnamefont {D.~A.}\ \bibnamefont {Williams}},
  \ and\ \bibinfo {author} {\bibfnamefont {H.}~\bibnamefont {Ahmed}},\ }\href
  {\doibase 10.1103/PhysRevB.50.8118} {\bibfield  {journal} {\bibinfo
  {journal} {Phys. Rev. B}\ }\textbf {\bibinfo {volume} {50}},\ \bibinfo
  {pages} {8118} (\bibinfo {year} {1994})}\BibitemShut {NoStop}%
\bibitem [{\citenamefont {Blonder}\ \emph {et~al.}(1982)\citenamefont
  {Blonder}, \citenamefont {Tinkham},\ and\ \citenamefont
  {Klapwijk}}]{Blonder1982}%
  \BibitemOpen
  \bibfield  {author} {\bibinfo {author} {\bibfnamefont {G.~E.}\ \bibnamefont
  {Blonder}}, \bibinfo {author} {\bibfnamefont {M.}~\bibnamefont {Tinkham}}, \
  and\ \bibinfo {author} {\bibfnamefont {T.~M.}\ \bibnamefont {Klapwijk}},\
  }\href@noop {} {\bibfield  {journal} {\bibinfo  {journal} {Phys. Rev. B.}\
  }\textbf {\bibinfo {volume} {25}},\ \bibinfo {pages} {4515} (\bibinfo {year}
  {1982})}\BibitemShut {NoStop}%
\bibitem [{\citenamefont {Octavio}\ \emph {et~al.}(1983)\citenamefont
  {Octavio}, \citenamefont {Tinkham}, \citenamefont {Blonder},\ and\
  \citenamefont {Klapwijk}}]{Octavio1983}%
  \BibitemOpen
  \bibfield  {author} {\bibinfo {author} {\bibfnamefont {M.}~\bibnamefont
  {Octavio}}, \bibinfo {author} {\bibfnamefont {M.}~\bibnamefont {Tinkham}},
  \bibinfo {author} {\bibfnamefont {G.~E.}\ \bibnamefont {Blonder}}, \ and\
  \bibinfo {author} {\bibfnamefont {T.~M.}\ \bibnamefont {Klapwijk}},\
  }\href@noop {} {\bibfield  {journal} {\bibinfo  {journal} {{Phys. Rev. B}}\
  }\textbf {\bibinfo {volume} {27}},\ \bibinfo {pages} {6739} (\bibinfo {year}
  {1983})}\BibitemShut {NoStop}%
\bibitem [{\citenamefont {Flensberg}\ \emph {et~al.}(1988)\citenamefont
  {Flensberg}, \citenamefont {Hansen},\ and\ \citenamefont
  {Octavio}}]{Flensberg1988}%
  \BibitemOpen
  \bibfield  {author} {\bibinfo {author} {\bibfnamefont {K.}~\bibnamefont
  {Flensberg}}, \bibinfo {author} {\bibfnamefont {J.~B.}\ \bibnamefont
  {Hansen}}, \ and\ \bibinfo {author} {\bibfnamefont {M.}~\bibnamefont
  {Octavio}},\ }\href@noop {} {\bibfield  {journal} {\bibinfo  {journal}
  {{Phys. Rev. B}}\ }\textbf {\bibinfo {volume} {38}},\ \bibinfo {pages} {8707}
  (\bibinfo {year} {1988})}\BibitemShut {NoStop}%
\bibitem [{\citenamefont {Kulik}(1970)}]{Kulik1970}%
  \BibitemOpen
  \bibfield  {author} {\bibinfo {author} {\bibfnamefont {I.~O.}\ \bibnamefont
  {Kulik}},\ }\href@noop {} {\bibfield  {journal} {\bibinfo  {journal} {Soviet
  Physics JETP}\ }\textbf {\bibinfo {volume} {30}},\ \bibinfo {pages} {944}
  (\bibinfo {year} {1970})}\BibitemShut {NoStop}%
\bibitem [{\citenamefont {Ishii}(1970)}]{Ishii1970}%
  \BibitemOpen
  \bibfield  {author} {\bibinfo {author} {\bibfnamefont {C.}~\bibnamefont
  {Ishii}},\ }\href@noop {} {\bibfield  {journal} {\bibinfo  {journal}
  {Progress of Theoretical Physics}\ }\textbf {\bibinfo {volume} {44}},\
  \bibinfo {pages} {1525} (\bibinfo {year} {1970})}\BibitemShut {NoStop}%
\bibitem [{\citenamefont {Bardeen}\ and\ \citenamefont
  {Johnson}(1972)}]{Bardeen1972}%
  \BibitemOpen
  \bibfield  {author} {\bibinfo {author} {\bibfnamefont {J.}~\bibnamefont
  {Bardeen}}\ and\ \bibinfo {author} {\bibfnamefont {J.~L.}\ \bibnamefont
  {Johnson}},\ }\href {\doibase 10.1103/PhysRevB.5.72} {\bibfield  {journal}
  {\bibinfo  {journal} {Phys. Rev. B}\ }\textbf {\bibinfo {volume} {5}},\
  \bibinfo {pages} {72} (\bibinfo {year} {1972})}\BibitemShut {NoStop}%
\bibitem [{\citenamefont {Likharev}(1979)}]{Likharev1979}%
  \BibitemOpen
  \bibfield  {author} {\bibinfo {author} {\bibfnamefont {K.~K.}\ \bibnamefont
  {Likharev}},\ }\href {\doibase 10.1103/RevModPhys.51.101} {\bibfield
  {journal} {\bibinfo  {journal} {Rev. Mod. Phys.}\ }\textbf {\bibinfo {volume}
  {51}},\ \bibinfo {pages} {101} (\bibinfo {year} {1979})}\BibitemShut
  {NoStop}%
\bibitem [{\citenamefont {Chrestin}\ \emph {et~al.}(1994)\citenamefont
  {Chrestin}, \citenamefont {Matsuyama},\ and\ \citenamefont
  {Merkt}}]{Chrestin1994}%
  \BibitemOpen
  \bibfield  {author} {\bibinfo {author} {\bibfnamefont {A.}~\bibnamefont
  {Chrestin}}, \bibinfo {author} {\bibfnamefont {T.}~\bibnamefont {Matsuyama}},
  \ and\ \bibinfo {author} {\bibfnamefont {U.}~\bibnamefont {Merkt}},\
  }\href@noop {} {\bibfield  {journal} {\bibinfo  {journal} {Phys. Rev. B}\
  }\textbf {\bibinfo {volume} {49}},\ \bibinfo {pages} {498} (\bibinfo {year}
  {1994})}\BibitemShut {NoStop}%
\bibitem [{\citenamefont {Dubos}\ \emph {et~al.}(2001)\citenamefont {Dubos},
  \citenamefont {Courtois}, \citenamefont {Pannetier}, \citenamefont {Wilhelm},
  \citenamefont {Zaikin},\ and\ \citenamefont {Sch{\"o}n}}]{Dubos2001}%
  \BibitemOpen
  \bibfield  {author} {\bibinfo {author} {\bibfnamefont {P.}~\bibnamefont
  {Dubos}}, \bibinfo {author} {\bibfnamefont {H.}~\bibnamefont {Courtois}},
  \bibinfo {author} {\bibfnamefont {B.}~\bibnamefont {Pannetier}}, \bibinfo
  {author} {\bibfnamefont {F.~K.}\ \bibnamefont {Wilhelm}}, \bibinfo {author}
  {\bibfnamefont {A.~D.}\ \bibnamefont {Zaikin}}, \ and\ \bibinfo {author}
  {\bibfnamefont {G.}~\bibnamefont {Sch{\"o}n}},\ }\href@noop {} {\bibfield
  {journal} {\bibinfo  {journal} {Phys. Rev. B}\ }\textbf {\bibinfo {volume}
  {63}},\ \bibinfo {pages} {064502} (\bibinfo {year} {2001})}\BibitemShut
  {NoStop}%
\bibitem [{\citenamefont {Giazotto}\ \emph {et~al.}(2004)\citenamefont
  {Giazotto}, \citenamefont {Grove-Rasmussen}, \citenamefont {Fazio},
  \citenamefont {Beltram}, \citenamefont {Linfield},\ and\ \citenamefont
  {Ritchie}}]{Giazotto2004}%
  \BibitemOpen
  \bibfield  {author} {\bibinfo {author} {\bibfnamefont {F.}~\bibnamefont
  {Giazotto}}, \bibinfo {author} {\bibfnamefont {K.}~\bibnamefont
  {Grove-Rasmussen}}, \bibinfo {author} {\bibfnamefont {R.}~\bibnamefont
  {Fazio}}, \bibinfo {author} {\bibfnamefont {F.}~\bibnamefont {Beltram}},
  \bibinfo {author} {\bibfnamefont {E.~H.}\ \bibnamefont {Linfield}}, \ and\
  \bibinfo {author} {\bibfnamefont {D.~A.}\ \bibnamefont {Ritchie}},\
  }\href@noop {} {\bibfield  {journal} {\bibinfo  {journal} {J. Supercond.}\
  }\textbf {\bibinfo {volume} {17}},\ \bibinfo {pages} {317} (\bibinfo {year}
  {2004})}\BibitemShut {NoStop}%
\bibitem [{Note1()}]{Note1}%
  \BibitemOpen
  \bibinfo {note} {The area of the two arms is obtained from the SEM image
  shown in Fig.~\ref {fig:1}(a). We have neglected a possible penetration of
  the magnetic field into the superconductor, because the London penetration
  depth of Nb is with 39 nm negligibly small \cite {Maxfield1965}.}\BibitemShut
  {Stop}%
\bibitem [{\citenamefont {Noguchi}\ \emph {et~al.}(1991)\citenamefont
  {Noguchi}, \citenamefont {Hirakawa},\ and\ \citenamefont
  {Ikoma}}]{Noguchi1991}%
  \BibitemOpen
  \bibfield  {author} {\bibinfo {author} {\bibfnamefont {M.}~\bibnamefont
  {Noguchi}}, \bibinfo {author} {\bibfnamefont {K.}~\bibnamefont {Hirakawa}}, \
  and\ \bibinfo {author} {\bibfnamefont {T.}~\bibnamefont {Ikoma}},\
  }\href@noop {} {\bibfield  {journal} {\bibinfo  {journal} {Phys. Rev. Lett.}\
  }\textbf {\bibinfo {volume} {66}},\ \bibinfo {pages} {2243} (\bibinfo {year}
  {1991})}\BibitemShut {NoStop}%
\bibitem [{\citenamefont {Olsson}\ \emph {et~al.}(1996)\citenamefont {Olsson},
  \citenamefont {Andersson}, \citenamefont {H\aa{}kansson}, \citenamefont
  {Kanski}, \citenamefont {Ilver},\ and\ \citenamefont
  {Karlsson}}]{Olsson1996}%
  \BibitemOpen
  \bibfield  {author} {\bibinfo {author} {\bibfnamefont {L.~O.}\ \bibnamefont
  {Olsson}}, \bibinfo {author} {\bibfnamefont {C.~B.~M.}\ \bibnamefont
  {Andersson}}, \bibinfo {author} {\bibfnamefont {M.~C.}\ \bibnamefont
  {H\aa{}kansson}}, \bibinfo {author} {\bibfnamefont {J.}~\bibnamefont
  {Kanski}}, \bibinfo {author} {\bibfnamefont {L.}~\bibnamefont {Ilver}}, \
  and\ \bibinfo {author} {\bibfnamefont {U.~O.}\ \bibnamefont {Karlsson}},\
  }\href@noop {} {\bibfield  {journal} {\bibinfo  {journal} {Phys. Rev. Lett.}\
  }\textbf {\bibinfo {volume} {76}},\ \bibinfo {pages} {3626} (\bibinfo {year}
  {1996})}\BibitemShut {NoStop}%
\bibitem [{\citenamefont {Capotondi}\ \emph {et~al.}(2005)\citenamefont
  {Capotondi}, \citenamefont {Biasiol}, \citenamefont {Ercolani},\ and\
  \citenamefont {Sorba}}]{Capotondi2005}%
  \BibitemOpen
  \bibfield  {author} {\bibinfo {author} {\bibfnamefont {F.}~\bibnamefont
  {Capotondi}}, \bibinfo {author} {\bibfnamefont {G.}~\bibnamefont {Biasiol}},
  \bibinfo {author} {\bibfnamefont {D.}~\bibnamefont {Ercolani}}, \ and\
  \bibinfo {author} {\bibfnamefont {L.}~\bibnamefont {Sorba}},\ }\href@noop {}
  {\bibfield  {journal} {\bibinfo  {journal} {J. Cryst. Growth.}\ }\textbf
  {\bibinfo {volume} {278}},\ \bibinfo {pages} {538} (\bibinfo {year}
  {2005})}\BibitemShut {NoStop}%
\bibitem [{\citenamefont {Maxfield}\ and\ \citenamefont
  {McLean}(1965)}]{Maxfield1965}%
  \BibitemOpen
  \bibfield  {author} {\bibinfo {author} {\bibfnamefont {B.~W.}\ \bibnamefont
  {Maxfield}}\ and\ \bibinfo {author} {\bibfnamefont {W.~L.}\ \bibnamefont
  {McLean}},\ }\href {\doibase 10.1103/PhysRev.139.A1515} {\bibfield  {journal}
  {\bibinfo  {journal} {Phys. Rev.}\ }\textbf {\bibinfo {volume} {139}},\
  \bibinfo {pages} {A1515} (\bibinfo {year} {1965})}\BibitemShut {NoStop}%
\end{thebibliography}%

\end{document}